\documentclass[final,5p,twocolumn,authoryear]{elsarticle}

\usepackage{graphicx}

\usepackage{amssymb}
\usepackage{amsthm}
\usepackage{amsmath}
\usepackage{pdflscape}

\usepackage[T1]{fontenc}

\usepackage{xcolor}

\newcounter{bla}

\newcommand{\JLM}[1]{#1}

\journal{Computer Physics Communications}

\begin{document}

\begin{frontmatter}



\title{On the application of Jacobian-free Riemann solvers for relativistic radiation magnetohydrodynamics under M1 closure}


\author[a,b]{Jose López-Miralles\corref{author}}
\author[a,c]{Jose María Martí}
\author[a,c]{Manel Perucho}

\cortext[author] {Corresponding author.\\\textit{E-mail address:} jose.lopez-miralles@uv.es}
\address[a]{Departament d’Astronomía i Astrofísica, Universitat de València, Dr. Moliner 50, 46100, Burjassot (València), Spain}
\address[b]{Aurora Technology for the European Space Agency, ESAC/ESA, Camino Bajo del Castillo s/n, Urb. Villafranca del Castillo, Villanueva de la Cañada, Madrid, Spain}
\address[c]{Observatori Astron\`omic, Universitat de Val\`encia, C/ Catedr\`atic Jos\'e Beltr\'an 2, 46980, Paterna, Val\`encia, Spain.}

\begin{abstract}
Radiative transfer plays a major role in high-energy astrophysics. In multiple scenarios and in a broad range of energy scales, the coupling between matter and radiation is essential to understand the interplay between theory, observations and numerical simulations. In this paper, we present a novel scheme for solving the equations of radiation relativistic magnetohydrodynamics within the parallel code \textsc{Lóstrego}. These equations, which are formulated taking successive moments of the Boltzmann radiative transfer equation, are solved under the gray-body approximation and the M1 closure using an IMEX time integration scheme. The main novelty of our scheme is that we introduce for the first time in the context of radiation magnetohydrodynamics a family of Jacobian-free Riemann solvers based on internal approximations to the Polynomial Viscosity Matrix, which were demonstrated to be robust and accurate for non-radiative applications. The robustness and the limitations of the new algorithms are tested by solving a collection of one-dimensional and multi-dimensional test problems, both in the free-streaming and in the diffusion radiation transport limits. Due to its stable performance, the applicability of the scheme presented in this paper to real astrophysical scenarios in high-energy astrophysics is promising. In future simulations, we expect to be able to explore the dynamical relevance of photon-matter interactions in the context of relativistic jets and accretion discs, from microquasars and AGN to gamma-ray bursts.
\end{abstract}


\end{frontmatter}




\section{Introduction}

Among the wide and complex variety of physical processes that govern high-energy astrophysics, radiative transfer plays a fundamental role in a broad range of energy scales. For example, the existence of a radiation field that coexists with plasma and (strong) electromagnetic fields is of great importance for modelling core-collapse supernovae \citep{obergaulinger18,burrows19}, gamma-ray bursts \citep{aloy06,meszaros06,rivera16,rivera17,rivera18}, tidal disruption events \citep[see e.g.,][]{krolik20}, the extreme environment around magnetars, pulsars \citep{becker09} or accretion disks around black holes \citep{thorne74,zanotti11}. It is also essential for understanding the dynamical evolution of black-hole and neutron star mergers \citep{hayashi22,radice22}, as well as the post-merger state \citep{foucart15,shibata21}. In many of these systems, matter and radiation are strongly coupled and thus their feedback effects are dynamically relevant.

For accretion disks around black holes, matter and photons can interact when the mass accretion rate is near or over the Eddington limit (i.e., supercritical accretion flows). This is the case of Seyfert galaxies \citep{ramos17}, but also for some microquasars in the high-luminous state, which have also been proposed to be at the origin of the emission of extragalactic Ultra Luminous X-ray Sources \citep[ULXs;][]{king01,mineshige11}. The high-mass X-ray binary SS433 is a fiducial example of this type of accretion \citep{begelman06,khabi16}. In this scenario, geometrically thick disks are supported by radiation pressure. On the other hand, the dominance of radiation pressure in optically thick, geometrically thin (i.e., standard) accretion disks could be unstable to thermal and viscous instabilities \citep{shakura76,takahashi11}. Moreover, radiation pressure force could play a role in the acceleration of relativistic jets and/or winds \citep{sikora96,ohsuga05,okuda05,ohsuga09,okuda09,takeuchi10,ohsuga11,sadowski15,sananda21}, while radiation drag could act against it \citep{beskin04}.

Due to all of these reasons, radiative transfer plays a major role in high-energy astrophysics, and it is essential for understanding the interplay between theory, observations and numerical simulations. However, solving the full Boltzmann radiative transfer equation is in general a very computationally expensive task, so full three-dimensional global solutions are prohibited even for modern parallel architectures. In numerical simulations, the most common approach to assess the problem of radiative transfer is by radiative post-processing with ray-tracing algorithms \citep[see e.g.,][]{fromm17,fuentes18,Fromm19,fuentes21}, specially for those scenarios where the interaction between matter and the photon field can be neglected. This means that the radiative output is calculated once the hydrodynamical simulation has finished, avoiding the complexity of the coupling problem. In this kind of methods, the photon field is modelled as a large number of narrow beams (i.e., rays) that propagates through the computational grid. Although these strategies are generally accurate for a big range of scenarios, they are also computationally expensive and do not account for the backreaction of radiation in the plasma dynamics. One alternative approach (followed in this paper) consists on taking successive moments of the Boltzmann radiative transfer equation. These methods are usually faster and accurate and account for the photon-matter coupling, but they can suffer for high levels of numerical diffusion in some particular situations.

For example, the Flux-Limited Diffusion (FLD) approximation \citep{levermore81} takes a zeroth-order momentum of the Boltzmann equation to evaluate the radiation energy density. Then, radiation fluxes are evaluated as the gradient of the radiation energy density without solving the first moment equation. Although some authors reported appropriate results for the radiation field using the FLD approximation in the optically-thick regime, it might lead to wrong solutions when the optical depth is close to or lower than unity \citep{ohsuga11}. Therefore, in order to have a robust method for both optically-thin and optically-thick regimes, both zeroth and first-order moments of the transport equation need to be considered. However, when solving more than one moment equation, a closure identity that relates the second moment of radiation (i.e., the radiation pressure tensor) with one of the lower order moments (e.g., the radiation energy density) is required, in the same way as the equation of state relates the primitive thermal variables of the hydrodynamical system. This equation has the form, $\tilde{P}_r^{ ij}=\tilde{D}_r^{ij}\tilde{E}_r$, where $\tilde{E}_r$ is the radiation energy, $\tilde{P}_r^{ij}$ is the radiation pressure tensor and $\tilde{D}_r^{ij}$ is the Eddington tensor\footnote{In this expression, tildes indicate that all quantities are defined in a frame of reference moving with the radiation field  (i.e., the comoving frame), and it is the notation adopted for the rest of the paper.}. To select the particular form of the Eddington tensor, there are two main strategies. The first one, which is also the simplest, consists on assumming that the radiation field is isotropic in the comoving frame, taking $\tilde{D}_r^{ij}=\delta^{ij}/3$ \citep{mihalas84}. This strategy, called the Eddington approximation, is only accurate for optically thick radiation transport. Other possibility, which is the one included in our code, is the M1 closure \citep{minerbo78,levermore84}. This method takes into account the possible spatial anisotropies of the photon field and gives accurate results for both optically-thin and optically-thick regimes. With all of these considerations, the equations of relativistic radiation magnetohydrodynamics (Rad-RMHD) can be written as a system of conservations laws, where  radiation-matter coupling appears as source terms in the equations. The main drawback of this approach is that, specially in the optically thick regime when matter and radiation interact more frequently, the equations of Rad-RMHD might become \textit{stiff}, meaning that the time scales of radiation processes (i.e., heating/cooling and scattering) might be too short compared to the dynamical scales of the plasma. We overcome this issue by using an implicit-explicit (IMEX) Runge-Kutta time integration method \citep{taka13}, where the spatial derivatives on the Rad-RMHD equations are treated explicitly (i.e., with the same methods used for non-radiation RMHD), while interaction terms that account for the exchange of energy and momentum between matter and radiation are integrated implicitly. 

In the context of Rad-RHD and Rad-RMHD, other authors have followed similar approaches. \cite{taka13} proposed an IMEX scheme for solving the equations of Rad-RHD taking zeroth and first moment equations of the radiative transfer equation. A similar approach was followed by \cite{taka132} to solve special relativistic, resistive radiation magnetohydrodynamics equations, consistently updated using the M1 closure. This same closure was included in the codes R-CAFE \citep{rivera19} and HARMRAD \citep{mckinney14} in the context of special Rad-RHD and general relativistic radiation magnetohydrodynamics, respectively. \cite{weih20} also followed a two-moment scheme within an IMEX scheme under the M1 closure for general-relativistic radiation hydrodynamics. \cite{pluto19} included an independent module of radiation transport in the freely available PLUTO code, using an IMEX approach under the M1 closure relation. Other authors, for example \cite{miniati07} or \cite{sekora09}, considered a slightly different method. They follow a higher order modified Godunov scheme that directly couples \textit{stiff} source term effects to the hyperbolic structure of the system of conservation laws. This method is composed of a predictor step based on Duhamel’s principle and a corrector step based on Picard iteration.


In this paper, we present a new scheme for treating radiation transport within the code \textsc{Lóstrego} \citep[][hereinafter, LM22]{miralles22}, by which we solve the Rad-RMHD equations using an IMEX Runge-Kutta time integration method under the M1 closure for the radiation field. However, what makes our approach unique among others is the introduction of a new family of Jacobian-free approximate Riemann solvers based on Polynomial Viscosity Matrix (PVM) methods \citep[][and references therein]{castro17}, which has never been applied before in the context of Rad-RMHD. For the sake of completeness, we have also introduced in \textsc{Lóstrego} a new family of high-order reconstruction methods \citep[Monotonicity Preserving, MP;][]{suresh97} and a five-step Runge-Kutta integration algorithm \citep{balsara01}. These high-order methods are properly tested in the numerical benchmark section (see Sec.~\ref{sec4}) with a classical test problem in relativistic magnetohydrodynamics.


The paper is organized as follows: in Sec.~\ref{sec2} we briefly describe the theoretical basis of radiative transfer and the system of equations of relativistic radiation magnetohydrodynamics. In Sec.~\ref{sec3} we describe our scheme and the new methods and algorithms included in \textsc{Lóstrego}. In Sec.~\ref{sec4} we provide a benchmark of one-dimensional and multi-dimensional test problems to demonstrate that the scheme is robust and stable in different radiative scenarios. In Sec.~\ref{sec5} we discuss the results of the benchmark and we summarize the main conclusions of the paper.

\section{Governing equations}
\label{sec2}

\subsection{Radiation RMHD}
\label{sec2.1}

The Rad-RMHD system of partial differential equations in the Minkowski metric\footnote{We assume a metric signature $(-,+,+,+)$. Greek subscripts in 4-vectors run from 0 to 3. Latin indices run from 1 to 3. In the following, we use a system of units where $c=1$ and a factor of $1/\sqrt{4\pi}$ is absorbed in the definition of the magnetic field.} and Cartesian coordinates can be written as a system of conservation laws \citep[see e.g.,][]{pluto19}:
\begin{equation}
    \partial_t\boldsymbol{U}+\partial_i\boldsymbol{F}^i=\boldsymbol{S},
    \label{eq0}
\end{equation}
where $\boldsymbol{U}=\{D,S^j,\tau_e,B^j,E_r,F_r^j\}$ is a vector of conserved variables, $D$ is the relativistic rest mass density, $S^j$ is the momentum density of the magnetized fluid, $\tau_e$ is the energy density (all of them measured in the Eulerian frame), and $\boldsymbol{F^i}$ are the vector of fluxes for each spatial direction. These two vectors can be expressed as a function of a set of primitives $\boldsymbol{V}=\{\rho,v^j,p,B^j,E_r,F_r^j\}$ through the following relations:
\begin{equation}
\label{eqU}
\boldsymbol{U}=
    \begin{pmatrix}
D\\ 
S^j\\ 
\tau_e\\ 
B^j\\ 
E_r\\
F_r^j\\

\end{pmatrix}=
    \begin{pmatrix}
\rho W \\ 
\rho h^* W^2v^j-b^0b^j\\ 
\rho h^* W^2-p^*-b^0b^0-\rho W\\ 
B^j\\ 
E_r\\
F_r^j\\
\end{pmatrix},
\end{equation}
\begin{equation}
\label{eqF}
\boldsymbol{F^i}=
        \begin{pmatrix}
\rho W v^i \\ 
\rho h^* W^2v^iv^j+p^*\delta^{ij}-b^ib^j\\ 
\rho h^* W^2v^i-b^0b^i-\rho W v^i\\ 
v^iB^j-B^iv^j\\ 
F_r^j\\
P_r^{ij}\\
\end{pmatrix},
\end{equation}
The set of primitive variables are the fluid rest-mass density $\rho$, fluid three-velocity $v_j$ , and gas pressure $p$. In Eqs. \ref{eqU} and \ref{eqF}, $p^*=p+|b|^2/2$ is the total pressure, $h^*=1+\epsilon+\gamma/\rho+|b|^2/\rho$ is the total specific enthalpy and $W=\sqrt{1+u_iu^i}$ is the Lorentz factor of the fluid, where $u^\mu=W~(1,v^i)$ is the relativistic four-velocity and $b^\mu$ is the relativistic magnetic field four-vector:
\begin{equation}
    b^0=W(\boldsymbol{v}\cdot\boldsymbol{B}),
\end{equation}
\begin{equation}
    b^i=\frac{B^i}{W}+v^ib^0.
\end{equation}
As in the case of RMHD, the magnetic field also obeys the divergence-free constraint:
\begin{equation}
    \nabla \cdot \boldsymbol{B} = 0.
\end{equation}
We shall consider a classical ideal gas with adiabatic exponent $\gamma$ which verifies the $\gamma$-law equation of state: 
\begin{equation}
    p=(\gamma-1)\rho\epsilon,
    \label{eos}
\end{equation}
where $\epsilon$ is the specific internal energy.
 
The magnetic field three-vector $B^j$, radiation energy density, $E_r$, and radiation flux, $F_r^j$, are both primitive and conserved variables. $P_r^{ij}$ is the radiation pressure tensor, which is related with the radiation fields by means of a specific closure relation, as described in Sec.~\ref{sec2.2}. The three radiation variables represent the first three moments of the radiation field and conform the radiation energy tensor in the laboratory frame \citep[see, e.g.,][]{mihalas84,taka13}:
\begin{equation}
    T_r^{\mu\nu}=
    \begin{pmatrix}
    E_r & F_r^i \\
    F_r^j & P_r^{ij}\\
    \end{pmatrix},
\end{equation}
which satisfies:
\begin{equation}
    \nabla_\mu T_r^{\mu\nu}=-\nabla_\mu(T_g^{\mu\nu}+T_{em}^{\mu\nu})=-G^{\nu},
    \label{G}
\end{equation}
where $G^{\nu}$ is the radiation four-force and $T_g^{\mu\nu}$, $T_{em}^{\mu\nu}$ and $T_r^{\mu\nu}$ are respectively the gas, electromagnetic and radiation components of the total energy-momentum tensor. The radiation four-force, $G^{\nu}$, represents the exchange of energy and momentum of matter and photons through absorption/emission and scattering processes, so the source terms in Eq.~\ref{eq0} are:
\begin{equation}
\boldsymbol{S}=
    \begin{pmatrix}
0\\ 
G^j\\ 
G^0\\ 
0\\ 
-G^0\\
-G^j\\
\end{pmatrix},
\label{sourceterms}
\end{equation}
which implies that the equations of evolution of relativistic density, $D$, and magnetic fields, $B^j$, are not affected by the photon field. The radiation four-force is explicitly given in the laboratory frame by \citep{taka13}:
\begin{equation}
    \begin{aligned}
    G^0 = -\rho\kappa~(4\pi \mathcal{B}W-WE_r+u_iF_r^i)-\\
    \rho\sigma_s\left[W(W^2-1)E_r+Wu_iu_jP_r^{ij}-(2W^2-1)~u_iF_r^i\right],
    \end{aligned}
    \label{g0}
\end{equation}

\begin{equation}
    \begin{aligned}
    G^i = -4\pi\rho\kappa \mathcal{B} u^i + \rho(\kappa+\sigma_s)(WF_r^i-u_jP_r^{ij})\\
    -\rho\sigma_su^i(W^2E_r-2Wu_jF_r^j+u_ju_kP_r^{jk}),
    \end{aligned}
    \label{gi}
\end{equation}
where $\kappa$ and $\sigma_s$ are respectively the frequency-averaged absorption and scattering coefficients measured in the comoving frame (i.e., the grey-body approximation). This means that Eq. \ref{G} is also a mixed-frame energy-momentum equation.

The variable $\mathcal{B}$ represents the fluid blackbody intensity, which is related with the comoving emissivity $\varepsilon$ through the Kirchhoff-Planck relation, $\varepsilon=\kappa \mathcal{B}$, which is valid for emission and absorption processes in matter in thermal equilibrium. Using the gas temperature, $\mathcal{B}$ can be determined as:
\begin{equation}
    \mathcal{B}=\frac{a_RT^4}{4\pi},
\end{equation}
where $a_R$ is a radiation constant (that is related with the Stefan-Boltzmann constant by $a_R=4\sigma_{\rm SB}$), and the gas temperature is given by the ideal gas equation of state:
\begin{equation}
    T=\frac{p \mu m_p}{\rho k_B},
\end{equation}
where $\mu$ is the mean molecular weight, $m_p$ is the proton mass and $k_B$ is the Boltzmann constant.

\subsection{Radiation closure}
\label{sec2.2}

Eq.~\ref{eos} provides a relation between the thermodynamic variables that closes the system of RMHD equations. For the Rad-RMHD system, an extra closure is needed to relate the radiation fields. This equation has the form:
\begin{equation}
    P_r^{ij}=D_r^{ij}E_r,
\end{equation}
where $D_r^{ij}$ is the Eddington tensor. Among the different types of closure relations proposed in the literature, the simplest approach is the Eddington approximation, where the Eddington tensor in the comoving frame takes the form $\tilde{D}_r^{ij}=\delta^{ij}/3$ \citep{mihalas84}. However, this method is only valid for the optically-thick radiation regime and it requires Lorentz transformations to calculate the radiation pressure tensor in the laboratory frame. Therefore, we have implemented the M1 closure \citep{minerbo78,levermore84}, which is valid in any reference frame and in both optical depth regimes, so it provides a better approximation to the radiation field than the Eddington approximation. Under this approach, the Eddington tensor is given by:
\begin{equation}
    D_r^{ij}=\frac{1-\xi}{2}\delta^{ij}+\frac{3\xi-1}{2}n^in^j,
\end{equation}
where $n^i=F_r^i/|\boldsymbol{F}_r|$ and $\xi$ is the Eddington factor:
\begin{equation}
    \xi=\frac{3+4f^2}{5+2\sqrt{4-3f^2}},
    \label{xi}
\end{equation}
where $f=|\boldsymbol{F}_r|/E_r$ is the reduced radiative flux. In the optically-thick regime, $|\boldsymbol{F}_r| \ll E_r$ and thus $f\rightarrow0$ which describes the diffusion limit, where $P_r^{ij}\approx(\delta^{ij}/3)~E_r$ (i.e., the Eddington approximation). In the optically-thin regime, $|\boldsymbol{F}_r| \approx  E_r$ and thus $f\rightarrow1$, which is associated to the free-streaming limit of the Rad-RMHD equations.

\section{Numerical methods}
\label{sec3}

Our new radiation scheme is implemented in \textsc{Lóstrego}, a 3D RMHD parallel code. Since the coupling between matter and radiation is represented by a collection of source terms (Eq.~\ref{sourceterms}), the general structure of the RMHD code can be preserved, making the extension to Rad-RMHD more natural. This means that all methods described in the appendix of LM22, as well as its parallelization strategy, are still valid and we do not reproduce the techniques in this paper. Instead, in this section we will concentrate on the methods and algorithms that are specific to solve the system of equations of Rad-RMHD.

\subsection{Numerical scheme}

\textsc{Lóstrego} is based on multidimensional High Resolution Shock-Capturing (HRSC) methods where we follow the finite volumes (FV) scheme and dimensional splitting, taking the integral form of the conservation laws and cell-averaged values. A conservative one-dimensional discretization of Eq.~\ref{eq0} yields:
\begin{equation}
    \boldsymbol{U}_{i}^{n+1}=\boldsymbol{U}_{i}^{n}-\frac{\Delta t}{\Delta x}\left(\boldsymbol{\hat{F}}_{i+1/2}-\boldsymbol{\hat{F}}_{i-1/2}\right)+\Delta t~\boldsymbol{S}_i^{n},
\label{full}
\end{equation}
where $\boldsymbol{U}_{i}^{n}$ and $\boldsymbol{S}_i^{n}$ are respectively the conserved variables and the source terms at  $t=n~\Delta t$, and $\boldsymbol{\hat{F}}_{i\pm 1/2}$ are the numerical fluxes. In Eq.~\ref{full}, $x=x_i$ represents the position of the cell center and $x=x_{i\pm 1/2}$ the position of the right and left cell interfaces, respectively. The element $\Delta x$ is the cell size and $\Delta t$, the time step. In the FV formalism, $\boldsymbol{U}_{i}^{n}$ and $\boldsymbol{S}_i^{n}$ represent an approximation to the averages of the corresponding quantities over the cell volume, while numerical fluxes $\boldsymbol{\hat{F}}_{i + 1/2}$ represent an approximation to the average of fluxes between $t=n~\Delta t$ and $t=(n+1)~\Delta t$ at cell interfaces, obtained by solving Riemann problems with initial data at $t=n~\Delta t$.

In our implementation of the IMEX scheme, we follow an extension of the total-variation-diminishing (TVD) Runge-Kutta schemes of \cite{shu89} that we implemented in \textsc{Lóstrego} to integrate the equations of RMHD (see LM22). In this case, the one-dimensional version of Eq.~\ref{eq0} is solved in two steps. First, the explicit step of the algorithm consists on solving the hyperbolic part of the equation:
\begin{equation}
    \boldsymbol{U}_{i}^{*}=\boldsymbol{U}_{i}^{n}-\frac{1}{\Delta x}\left(\boldsymbol{\hat{F}}_{i+1/2}-\boldsymbol{\hat{F}}_{i-1/2}\right),
\label{explicit}
\end{equation}
where $\boldsymbol{U}_{i}^{*}$ are the conserved variables after the explicit step. The new Riemann solvers, which are the central element of the explicit step, are described in Sec.~\ref{riemann}. Secondly, after every flux integration, the equation that gives the solution at $t=(n+1)~\Delta t$,
\begin{equation}
    \boldsymbol{U}_{i}^{n+1}=\boldsymbol{U}_{i}^{*}+\Delta t~\boldsymbol{S}_i^n,
\label{impliciteq}
\end{equation}
is solved implicitly as described in Sec.~\ref{implicit}.
For example, the third-order Runge-Kutta method (RK3) of \cite{shu89} is adapted to this purpose as:
\begin{equation}
\begin{aligned}
\boldsymbol{U}_i^{*}=\boldsymbol{U}_i^{(0)}+\Delta t~L\left(\boldsymbol{U}_i^{(0)}\right)\\
\boldsymbol{U}_i^{(1)}=\boldsymbol{U}_i^{*}+\Delta t~\boldsymbol{S}_i^{(0)}\\
\boldsymbol{U}_i^{*}=\frac{3}{4}\boldsymbol{U}_i^{(0)}+\frac{1}{4}\boldsymbol{U}_i^{(1)}+\frac{1}{4}\Delta t~ L\left(\boldsymbol{U}_i^{(1)}\right)\\
\boldsymbol{U}_i^{(2)}=\boldsymbol{U}_i^{*}+\frac{1}{4}\Delta t~\boldsymbol{S}_i^{(1)}\\
\boldsymbol{U}_i^{*}=\frac{1}{3}\boldsymbol{U}_i^{(0)}+\frac{2}{3}\boldsymbol{U}_i^{(2)}+\frac{2}{3}\Delta t~L\left(\boldsymbol{U}_i^{(2)}\right)\\
\boldsymbol{U}_i^{(3)}=\boldsymbol{U}_i^{*}+\frac{2}{3}\Delta t~\boldsymbol{S}_i^{(2)},\\
\end{aligned}
\end{equation}
where $L\left(\boldsymbol{U}_i^{(n)}\right)$ is the upwind differencing operator:
\begin{equation}
    L\left(\boldsymbol{U}_i^{(n)}\right)=-\frac{1}{\Delta x}\left(\boldsymbol{\hat{F}}_{i+1/2}-\boldsymbol{\hat{F}}_{i-1/2}\right).
\end{equation}

\subsection{Explicit step: the Riemann problem}
\label{riemann}

The explicit step of the algorithm is based on the reconstruct-solve-update strategy. First, cell-average primitive variables are reconstructed to the cell interfaces by means of one of the following piecewise linear methods (PLM): MinMod \citep{roe86}, MC \citep{VanLeer77} or VanLeer \citep{vanleer74}. The slope limiters implemented in these methods are essential to preserve the monotonicity and TV-stability of the algorithm. For the sake of completeness, we have also included the high-order MP reconstruction of \cite{suresh97}, although for the purposes of this work we restrict its applicability to problems with no radiative transport. In RMHD, we reduce the reconstruction to zero-th order (i.e., Godunov reconstruction) when the algorithm lead to non-physical solutions, as it is the case for superluminical velocities, $|\boldsymbol{v}|^2>1$. In Rad-RMHD, we also need to provide an upper limit for the radiation flux, such that:
\begin{equation}
    |\boldsymbol{F}_r| \leq E_r.
\end{equation}
After reconstructing the primitives at the cell boundaries, we solve an initial value problem by using an approximate Riemann solver to obtain the numerical fluxes at each cell interface. Finally, once fluxes are known at each cell face, the conserved variables are evolved explicitly in time according to Eq.~\ref{explicit}. Although the HLL-family of Riemann solvers can be naturally extended to Rad-RMHD, matter and radiation usually have different maximum/minimum characteristic velocities. We devote the rest of this section to describe the Rad-RMHD version of the HLL Riemann solver, what type of limitations one can expect and a different approach based on a new family of algorithms. 

\subsubsection{Radiative HLL}

A radiative version of the HLL Riemann solver was proposed, for instance, by \cite{gonzalez07} and \cite{pluto19}. In classical fluid dynamics \citep{harten83,toro13}, the initial discontinuity is decomposed into two fast magnetosonic waves with characteristic speeds $\lambda_L, \lambda_R$, such that the internal fluxes are derived from the Rankine-Hugoniot jump conditions across the two magnetosonic waves. The numerical fluxes at $x=x_{i+1/2}$ are then given by:
\begin{equation}
\boldsymbol{\hat{F}}_{i+1/2}=\left\{\begin{matrix}
\boldsymbol{F}_L,  & \text{if}\hspace{0.1cm} \lambda_L>0, \\ 
\hspace{0.4cm}\boldsymbol{F}^{\rm hll}_{i+1/2},  & \hspace{0.78cm}\text{if}\hspace{0.1cm} \lambda_L\le 0 \le \lambda_R, \\
\boldsymbol{F}_R, & \text{if}\hspace{0.1cm} \lambda_R<0. 
\end{matrix}\right.
\end{equation}
where:
\begin{equation}
   \boldsymbol{F}^{\rm hll}_{i+1/2}=\frac{\lambda_R \boldsymbol{F}_L - \lambda_L \boldsymbol{F}_R + \lambda_R \lambda_L(\boldsymbol{U}_R - \boldsymbol{U}_L)}{\lambda_R-\lambda_L},
\end{equation}
with $\boldsymbol{U}_{L, \, R} = \boldsymbol{U}^n_{i, \, i+1}$ and $\boldsymbol{F}_{L, \, R} = \boldsymbol{F}(\boldsymbol{U}_{L, \, R})$.
The new version is based on the fact that the Jacobian matrix for the radiation part of Eq.~\ref{eq0}, $J^i_{\rm rad}$, is only a function of the radiation energy density $E_r$ and the radiation flux $\boldsymbol{F}_r$:
\begin{equation}
  J^i_{\rm rad}=\begin{pmatrix}
 \partial F_r^i/\partial E_r&\partial F_r^i/\partial F_r^j \\ 
 \partial P_r^{ij}/\partial E_r&\partial P_r^{ij}/\partial F_r^j 
\end{pmatrix},  
\label{jacobian}
\end{equation}
so the Jacobian of the system of Rad-RMD equations can be decomposed in two different blocks: one submatrix for the magnetofluid and other submatrix for the radiation field. This means that the wave speeds $\lambda_L,\lambda_R$ can be calculated independently for each of these blocks, avoiding the excessive numerical diffusion that appears when employing the same velocities for both subsystems. On the one hand, for the RMHD block, these characteristic speeds (which are the maximum and minimum eigenvalues of the Jacobian matrix of the reconstructed states) are the solution of a quartic equation, for which we employ a root-finding method \citep{anile89,anton10}. On the other hand, for the radiation block, the full set of eigenvalues are given by the following analytical expressions \citep{audit02,skinner13,pluto19}:
\begin{equation}
    \lambda_{r1}=\frac{f\cos{\theta}-\zeta(f,\theta)}{\sqrt{4-3f^2}},
    \label{lam1}
\end{equation}
\begin{equation}
    \lambda_{r2}=\cos{\theta}\frac{3\xi(f)-1}{2f},
\end{equation}
\begin{equation}
      \lambda_{r3}=\frac{f\cos{\theta}+\zeta(f,\theta)}{\sqrt{4-3f^2}},  
      \label{lam2}
\end{equation}
where $\xi(f)$ is given by Eq.~\ref{xi} and $\zeta(f,\theta)$ by:
\begin{equation}
\begin{aligned}
\zeta(f,\theta)=\Biggl[\frac{2}{3}\left(4-3f^2-\sqrt{4-3f^2}\right)+\\
2\cos{\theta}^2\left(2-f^2-\sqrt{4-3f^2}\right)\Biggr]^{1/2},    
\end{aligned}
\label{lam3}
\end{equation}
where $\theta$ represents the angle between $\boldsymbol{F}_r$ and the upwind direction $\boldsymbol{\hat{e}}_d$, adopting the same notation as in \cite{pluto19}. A different approach is followed by \cite{taka13}, where the wave velocities were computed from the Jacobian matrix and tabulated before time integration. In the free-streaming limit (i.e., $f\rightarrow1$), the three eigenvalues are degenerate and become $\lambda_{r1}=\lambda_{r2}=\lambda_{r3}\rightarrow\cos{\theta}$, such that in the parallel direction to $\boldsymbol{F}_r$ the speed of light is recovered. In the same way, in the perpendicular direction, there is no transport of radiation. On the other hand, in the diffusion limit (i.e., $f\rightarrow0$), the three characteristic speeds are $\lambda_{r2}\rightarrow 0$ and $\lambda_{r1,r3}\rightarrow\pm 1/\sqrt{3}$. However, for optically thick media, these speeds can be overestimated, leading to excessive numerical diffusion. To avoid it, \cite{sadowski13} suggests to locally limit the maximum and minimum speeds by means of:
\begin{equation}
    \lambda_{r,L}=\text{max}\left(\lambda_{r1},-\frac{4}{3\tau}\right)\\
    \lambda_{r,R}=\text{min}\left(\lambda_{r3},+\frac{4}{3\tau}\right),
    \label{limits}
\end{equation}
where $\tau=\rho W(\kappa+\sigma_s)\Delta x$ is the optical depth along one single cell.

Although HLL is usually a robust solver, it can be rather diffusive for some applications. Thus, to improve the accuracy of the solutions, \cite{pluto19} presented a novel version of the three-wave HLLC Riemann solver of \cite{mignone05} and \cite{Mignone06} for the radiation transport. However, since the outermost velocities must be limited in optically thick cells, their scheme is no longer valid and the algorithm must be switched to the standard HLL when this occurs. This means that the radiative HLLC solver of \cite{pluto19} can only improve the accuracy of Rad-HLL in optically thin media. \JLM{Nevertheless, to the best of our knowledge, the five-wave HLLD scheme of \cite{mignone09}, which is also included in \textsc{Lóstrego} (see LM22), has not been adapted for radiation transport.}

\subsubsection{Polynomial Viscosity Matrix (PVM) Riemann solver}
\label{PVM}

As discussed in the previous section, one fundamental obstacle to solve the system of Rad-RMHD equations is that the characteristic speeds of the two individual blocks can be, in general, very different. This means that, using the same signals for both systems, could lead to unacceptable levels of numerical diffusion. Moreover, in the optically thick regime, the maximum and minimum characteristic speeds have to be limited by Eq.~\ref{limits}, what makes that Rad-HLLC only improves the accuracy of the standard HLL Riemann solver in the optically thin limit. Thus, Jacobian-free low dissipative solvers are highly desirable for these reasons.

The \JLM{Polynomial} Viscosity Matrix (PVM) methods were introduced by \cite{castro12} for hyperbolic systems. These solvers are defined in terms of viscosity matrices that are based on polynomial evaluations of a Roe matrix \citep[see e.g.,][]{cargo97,toro13} or the Jacobian of the flux at some average value. \cite{castro14} further extended the idea to the case of Rational Viscosity Matrix functions (RVM), while \cite{castro16} applied the technique to the Dumber-Osher-Toro (DOT) schemes, which are a simpler approximation of the classical Osher-Solomon method \citep{osher82,castro16}. Later, \cite{castro17} investigated two Jacobian-free implementations of the PVM/RVM methods in the context of the special RMHD equations, one based on Chebyshev polynomials and other built using internal approximations to the absolute value function. All together, these methods form a new family of approximate Riemann solvers. 

In the rest of this section, we will concentrate on PVM solvers based on internal approximations of n-degree (hereinafter, PVM-int-n). The main advantage of these techniques is that they guarantee the stability conditions needed to ensure the robustness and convergence of the algorithm. Moreover, the underlying polynomials allows a Jacobian-free formulation of the scheme, where only evaluations of the flux and vector operations are involved. This is particularly relevant to solve the system of Rad-RMHD equations, where the eigenvalues of the radiation Jacobian matrix have to be calculated independently of the magnetofluid subsystem.

In general, the numerical fluxes of any hyperbolic system of conserved equations have the following form:
\begin{equation}
    \boldsymbol{\hat{F}}_{i+1/2}=\frac{\boldsymbol{F}_L + \boldsymbol{F}_R}{2}-\frac{1}{2}Q_{i+1/2}~(\boldsymbol{U}_R - \boldsymbol{U}_L),
    \label{generalflux}
\end{equation}
where $Q_{i+1/2}$ is the numerical viscosity matrix, and $\boldsymbol{U}_{L, \, R} = \boldsymbol{U}^n_{i, \, i+1}$, $\boldsymbol{F}_{L, \, R} = \boldsymbol{F}(\boldsymbol{U}_{L, \, R})$. In the Roe's method, the viscosity matrix can be written as $Q_{i+1/2}=|A_{i+1/2}|$, where $A_{i+1/2}$ is a Roe matrix of the system \citep{torrilhon12,cordier14}. The PVM method allows to approximate the viscosity matrix by: 
\begin{equation}
    Q_{i+1/2}=|\lambda_{i+1/2,\text{max}}|~p(|\lambda_{i+1/2,\text{max}}|^{-1}~A_{i+1/2}),
\end{equation}
where $p(x)$ is a polynomial approximation of $|x|$ in the interval $[-1,1]$ and $\lambda_{i+1/2,\text{max}}$ is an upper bound to the maximum eigenvalue of the Roe matrix. In the internal polynomial approximation, $p(x)$ is iteratively constructed as:
\JLM{
\begin{equation}
    p_0(x)=1,\\
    p_{k}(x)=\frac{1}{2}~(2p_{k-1}(x)-p_{k-1}(x)^2+x^2),
    \label{iterat}
\end{equation}
}
where \JLM{$k=1,2...$} determines the degree of the approximation as \JLM{$\text{deg}(p_k)=2^k$}. Following \cite{castro17}, we consider \JLM{$k=3$} ($\text{deg}(p_3)=8$) and we assume $\lambda_{i+1/2,\text{max}}=1$, which is a reasonable upper bound for radiation transport. Using the explicit form, Eq.~\ref{iterat} becomes:
\begin{equation}
    p_3(x)=x^2(x^2(x^2(\alpha_0 x^2+\alpha_1)+\alpha_2)+\alpha_3)+\alpha_4,
\end{equation}
with coefficients $\alpha_0=-1/128$, $\alpha_1=3/32$, $\alpha_2=-23/64$, $\alpha_3=31/32$, $\alpha_4=39/128$ \citep{castro17}. With all of these ingredients, the term $Q_{i+1/2}~(\boldsymbol{U}_{R} - \boldsymbol{U}_L)$ in Eq.~\ref{generalflux} can be computed for each cell interface using the Horner's algorithm \citep{horner19}. Our implementation is summarized as follows for a polynomial of \JLM{$\text{deg}(p_k)=2^k$}:
\begin{enumerate}

    \item Starting from the reconstructed primitive variables $\boldsymbol{V}_L, \boldsymbol{V}_R$, calculate $\boldsymbol{U}_L, \boldsymbol{U}_R$ with Eq.~\ref{eqU} and $\boldsymbol{F}_L, \boldsymbol{F}_R$ with Eq.~\ref{eqF}. Define $\boldsymbol{U}_0=\boldsymbol{U}_R-\boldsymbol{U}_L$.\\
    
    \item Take $\boldsymbol{U}_m=(\boldsymbol{U}_L+\boldsymbol{U}_R)/2$ as an arbitrary state and evaluate the Jacobian matrix, $A\equiv A(\boldsymbol{U}_m)$, on this state. Recover the primitive variables for the arbitrary state, $\boldsymbol{V}_m$, by the inversion algorithm used, for example, in LM22.\\
    
    \item Define a new set of variables $\boldsymbol{U}_{\epsilon}=\boldsymbol{U}_m+\epsilon \boldsymbol{U}_0$, where $\epsilon=10^{-8}$, and recover $\boldsymbol{V}_{\epsilon}$. Calculate $\boldsymbol{F}(\boldsymbol{U}_m(\boldsymbol{V}_m))$ and $\boldsymbol{F}(\boldsymbol{U}_{\epsilon}(\boldsymbol{V}_{\epsilon}))$ with Eq.~\ref{eqF}.\\
    
    \item Define a new set of variables $\tilde{\boldsymbol{U}}_{\epsilon}=\boldsymbol{U}_m+\boldsymbol{F}(\boldsymbol{U}_{\epsilon})-\boldsymbol{F}(\boldsymbol{U}_m)$ and recover $\tilde{\boldsymbol{V}}_{\epsilon}$.\\
    
    \item Calculate $\boldsymbol{U}_1=\alpha_0\tilde{\boldsymbol{U}}_0+\alpha_1\boldsymbol{U}_0$, where:
    \begin{equation}
    \begin{aligned}
        \tilde{\boldsymbol{U}}_0=A^2\boldsymbol{U}_0\equiv\\
        \Phi_{\epsilon}(\boldsymbol{U}_m(\boldsymbol{V}_m);\boldsymbol{U}_0(\boldsymbol{V}_0))\approx\frac{\boldsymbol{F}(\tilde{\boldsymbol{U}}_{\epsilon}(\tilde{\boldsymbol{V}}_{\epsilon}))-\boldsymbol{F}(\boldsymbol{U}_m(\boldsymbol{V}_m))}{\epsilon}.
        \label{utilde}
    \end{aligned}
    \end{equation}
    
    \item For \JLM{$l=1,2,...k$}, repeat steps 3 and 4 for \JLM{$\epsilon \boldsymbol{U}_0\rightarrow\epsilon \boldsymbol{U}_l$}. Calculate \JLM{$\boldsymbol{U}_{l+1}=\tilde{\boldsymbol{U}}_l+\alpha_{l+1} \boldsymbol{U}_0$}, where \JLM{$\tilde{\boldsymbol{U}}_l\equiv\Phi_{\epsilon}(\boldsymbol{U}_m(\boldsymbol{V}_m);\boldsymbol{U}_l(\boldsymbol{V}_l))$} is given by Eq.~\ref{utilde}.\\
    
    \item Finally, \JLM{$|A(\boldsymbol{U}_m)|\boldsymbol{U}_0\equiv Q_{i+1/2}(\boldsymbol{U}_R-\boldsymbol{U}_L)\approx \boldsymbol{U}_{k+1}$}.\\

    Nevertheless, we must stress the fact that in the optically-thick regime, this scheme might introduce undesirable levels of numerical diffusion (see details on Sec.~\ref{sec:pulse}). This pathology has been also identified with other types of Riemann solvers \citep[see e.g.,][]{gonzalez07}. Intuitively, one might be tempted to suppress the viscosity matrix when 
    the opacity is larger than 1. 

    However, since this yields oscillatory solutions, we propose to reduce the numerical viscosity, $Q_{i+1/2}(\boldsymbol{U}_R-\boldsymbol{U}_L)$, in optically-thick media in a smooth way inversely proportional to the cell local opacity, $\tau$, such that:
    \JLM{
    \begin{equation}
        \boldsymbol{\hat{F}}_{i+1/2}\approx\frac{\boldsymbol{F}_L +\boldsymbol{F}_R}{2}-\beta\frac{\boldsymbol{U}_{k+1}}{2},
    \end{equation}    
    }
    with 
    \begin{equation}
    \beta = {\rm min}\left\{1/\tau_L, 1/\tau_R, 1\right\}
    \label{beta}
    \end{equation}
    ($\tau_{L, \, R}$ stand for the cell opacities at the left and right, respectively, of interface $i+1/2$).

    This particular choice of the coefficient $\beta$ is based on the adimensional nature of the cell opacity and the robustness of the algorithm in the optically-thick regime.
\end{enumerate}

\subsection{Implicit step}
\label{implicit}

The implicit step of the algorithm consists of a series of steps to integrate Eq.~\ref{impliciteq} in time. Due to the particular form of the source terms (Eq.~\ref{sourceterms}), we only need to deal with the following reduced subsystem:
\begin{equation}
    \boldsymbol{U}_{r}^{n+1}=\boldsymbol{U}_{r}^{*}+\Delta t~\boldsymbol{S}_r^n
\label{timeadv}
\end{equation}
where $\boldsymbol{U}_r=(E_r,\boldsymbol{F}_r)$, while $\boldsymbol{S}_r=(-G^0,-G^i)$ is given by Eqs.~\ref{g0} and Eq.~\ref{gi}. This is because the total energy and the total momentum of the system:
\begin{equation}
    \tau_{e,t}=\tau_e^*+E_r^*,
    \label{totale}
\end{equation}
\begin{equation}
    \boldsymbol{S}_t=\boldsymbol{S}^*+\boldsymbol{F}_r^*,
    \label{totalm}
\end{equation}
 must be conserved during the implicit step, such that the energy density and momentum of the magnetofluid can be finally updated as $\tau_e^{n+1}=\tau_{e,t}-E_r^{n+1}$ and $\boldsymbol{S}^{n+1}=\boldsymbol{S}_t-\boldsymbol{F}_r^{n+1}$. This implicit step is based on an iterative process which comprises the following substeps:
 \begin{enumerate}
     \item Starting from the output of the explicit step $\boldsymbol{U}^*$, calculate the total energy and total momentum of the magnetofluid with Eq.~\ref{totale} and Eq.~\ref{totalm}.\\
     
     \item Construct the matrix $\boldsymbol{C}^{(m)}$, whose elements are \citep[see e.g.,][]{taka132}:
     \begin{equation}
         \begin{aligned}
         C^{(m)}_{11}=1-\Delta t~\rho W\left[-\kappa+\sigma_s\left(W^2-1+u_iu_jD_r^{ij}\right)\right]\\
         C^{(m)}_{1j+1}=-\Delta t~\rho u_j\left[\kappa-\sigma_s\left(2W^2-1\right)\right]\\
         C^{(m)}_{i+1\, 1}=-\Delta t~\rho\left[(\kappa+\sigma_s)u_jD^{ij}+\sigma_su^i\left(W^2+u_ku_lD_r^{kl}\right)\right]\\
         C^{(m)}_{i+1j+1}=\delta_j^{i}-\Delta t\rho W\left[(\kappa+\sigma_s)\delta_j^i+2\sigma_su^iu_j\right].
         \end{aligned}
     \end{equation}

\item Solve the system:
\begin{equation}
\boldsymbol{C}^{(m)}\begin{pmatrix}
E_r^{(m+1)}\\
F_r^{i,(m+1)}
\end{pmatrix}=\begin{pmatrix}
E_r^*+(4\pi\rho W\kappa B)^{(m)}~\Delta t\\
F_r^{i,*}+(4\pi\rho u^i\kappa B)^{(m)}~\Delta t
\end{pmatrix}    
\end{equation}
using an inversion method, for example, the LU-decomposition \citep{taka13}. For the first iteration, we take $E_r^{0}=E_r^*$ and $\boldsymbol{F}_r^{0}=\boldsymbol{F}_r^*$.\\
     \item Update the fluid energy density, $\tau_e^{(m+1)}=\tau_{e,t}-E_r^{(m+1)}$, and momentum, $\boldsymbol{S}^{(m+1)}=\boldsymbol{S}_t-\boldsymbol{F}_r^{(m+1)}$. For the relativistic rest-mass density and magnetic fields, we already have $D^{(n+1)}=D^*$ and $\boldsymbol{B}^{(n+1)}=\boldsymbol{B}^*$.\\
     \item Recover primitive variables with the inversion scheme.\\
     \item Check if the relative error of the radiation fields falls below a specified threshold; if not, update radiation fields and start a new iteration (substep 2).
 \end{enumerate}
 
 \section{Testing benchmark}
 \label{sec4}

In this section, we provide a collection of one dimensional and multidimensional numerical problems. Unless otherwise stated, all tests were performed using the PVM-int-8 Riemann solver and the second-order piecewise linear VanLeer reconstruction algorithm. In multidimensional problems, the VanLeer reconstruction algorithm is degraded to the more diffusive MinMod slope limiter around strong shocks \citep{mignone05}. For time integration, we used the IMEX third-order TVD Runge-Kutta algorithm \citep{shu89} with CFL=0.3. Magnetic field divergence free constraint is preserved by using the CT method, where electromotive forces were averaged according to the CT-contact formalism \citep{gardiner05}. In high magnetization regimes, the relativistic correction scheme of \cite{marti152} was used to correct the conserved variables after each time integration.

\subsection{One dimensional tests}

We propose a collection of 1D problems to test the performance of the new methods in the context of RMHD and Rad-RMHD. First, we solve a large-amplitude circularly-polarized Alfvén wave without radiation to demonstrate that the code achieves third order and fifth order of accuracy using high-order spatial reconstruction together with third order and fourth order Runge Kutta, respectively. Secondly, we test the explicit part of the new radiation module with a Riemann problem for the optically-thin radiation transport, neglecting any interaction between matter and radiation. Finally, we implemented a collection of shock problems and the diffusion of a radiation pulse without neglecting the interaction terms, in order to test the performance of the new IMEX scheme.

\subsubsection{Circularly-polarized Alfvén wave}
\label{dzaw}

We considered a smooth RMHD test problem that consists on the propagation of a large-amplitude, circularly polarized Alfvén wave along a uniform background magnetic field $B_0$. This test was previously used by \cite{delzanna07} and \cite{beckwith11} to test the accuracy of their codes ECHO and ATHENA, respectively. \cite{marti15} also reported second-order accuracy for their code with this test. The transverse components of the velocity field were initialized as:
\begin{equation}
    v_y = -A\cos{\left(\frac{2\pi}{\lambda}(x-v_at)\right)}, v_z = -A\sin{\left(\frac{2\pi}{\lambda}(x-v_at)\right)},
\end{equation}
where $A$ is the amplitude of the wave, and $\lambda$ its wavelength, and
\begin{equation}
    B_x=B_0, B_y=-B_0v_y/v_a, B_z=-B_0v_z/v_a.
\end{equation}
In the previous expressions, the speed of the Alfv\'en wave, $v_a$, is given by:
\begin{equation}
    v_a=\pm\sqrt{\frac{B_0^2~(1-A^2)}{\rho_0h_0+B_0^2~(1-A^2)}},
\end{equation}
where $\rho_0$ and $h_0$ are the density and the specific enthalpy of the background uniform medium. For this, we choose $\rho_0 = 1$, $h_0 = 5$ ($\gamma = 4/3$) and $B_0 = 1$. The amplitude of the wave is taken as:
\begin{equation}
    A=\sqrt{\frac{2}{7+3\sqrt{5}}}.
\end{equation}
To test the order of our code with this problem, we run a collection of simulations in the domain $[0,2\pi]$ for different spatial resolutions (i.e, from N=16 to N=512 zones) and we measured the L1-errors in one of the transverse components of the velocity field (i.e., $v_z$) after one entire wave period ($T=\lambda/v_a$), comparing with the initial conditions. 
This process was repeated for all the Riemann solvers (HLL, HLLC, HLLD) and all the piecewise linear methods (MinMod, MC, VanLeer) already available in \textsc{Lóstrego}. Additionally, we also tested the MP reconstruction of \cite{suresh97} together with HLLD and the new PVM-int-8 Riemann solver. To highlight the performance of the solver, we have also run the test with the first-order version of the code (i.e, Godunov spatial reconstruction).

\begin{figure}
	\includegraphics[width=\linewidth]{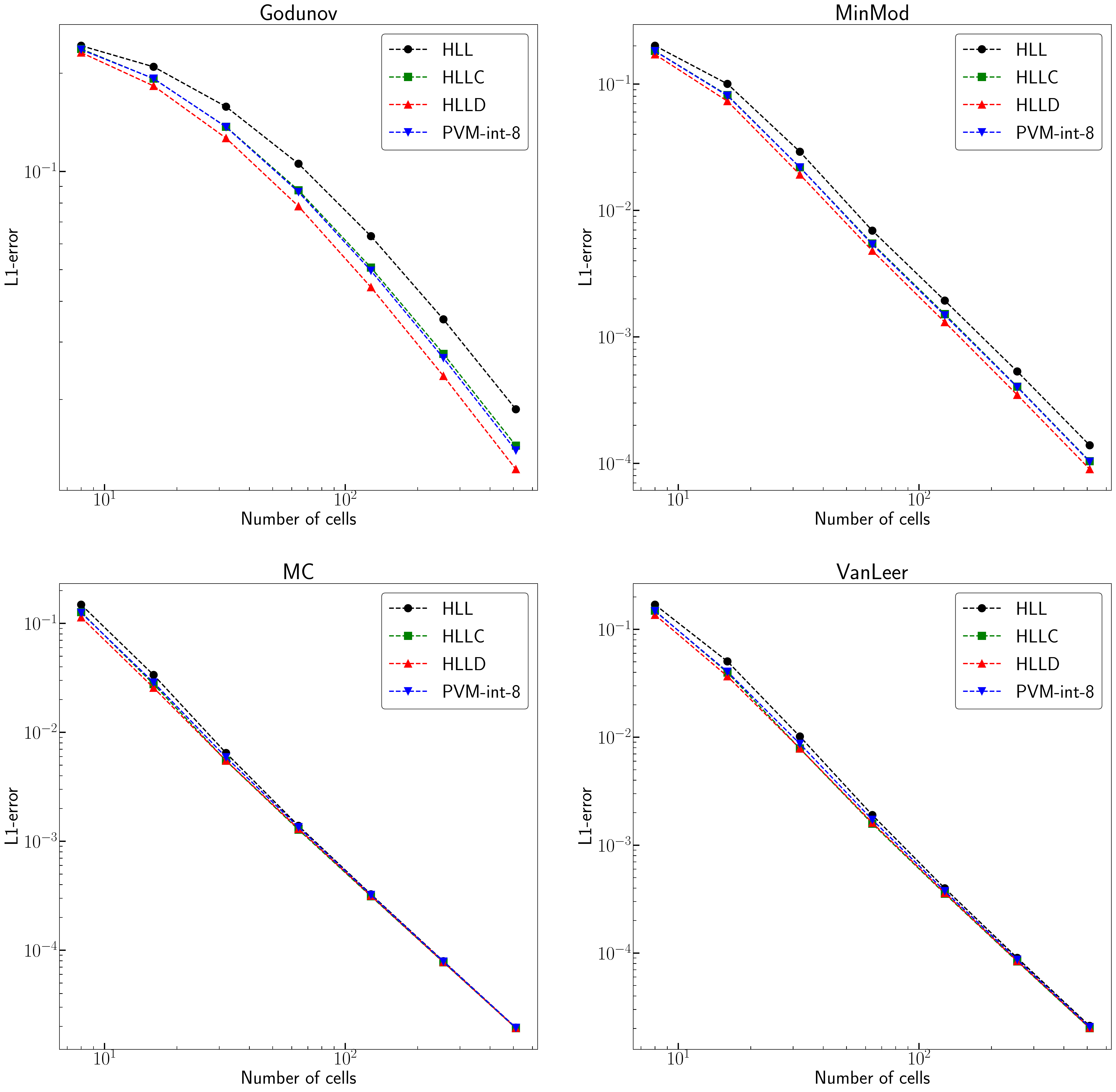}
    \caption{L1 errors of $v_z$ as a function of the number of cells for different Riemann solvers and spatial zeroth and first order reconstruction techniques.}
  \label{L1}
\end{figure}

The results of our analysis are shown in Tables \ref{table1}-\ref{table4} and in Fig.~\ref{L1}. We demonstrated that our code achieves first-order accuracy in the case of Godunov reconstruction and second-order accuracy for all the PLM methods, although in the latter case MinMod generally yields higher errors than its counterparts. In the case of HLLC (Table \ref{table2}), our results are the same as those obtained in \cite{marti15}. HLLD (Table \ref{table3}) shows similar performance than HLLC, as it is expected for smooth solutions. Using MP3 reconstruction together with HLLD and RK3, we achieved third order of accuracy, but we were not able to increase the order of the code using fifth-order spatial reconstruction (i.e., MP5).

This means that, for smooth RMHD problems, RK3 limits the accuracy of the code to third order, even should we employ higher-order spatial reconstructions. This is not the case, however, if we use a five-step RK4 algorithm \citep[][and references therein]{balsara17}, with which the code achieves the nominal fifth order of accuracy (MP5+RK4 in Table \ref{table3}). Finally, the results of the PVM-int-8 solver shown in Table \ref{table4} are very similar to those found with HLLC (see also Fig.~\ref{L1}). We also proved third order of accuracy with this new Riemann solver together with the MP3 reconstruction and the RK3 Runge-Kutta algorithm.


\begin{table*}
\centering
\caption{Code accuracy for the large-amplitude circularly polarized Alfvén wave test employing the HLL Riemann solver and zeroth/first-order reconstruction. The third-order Runka-Kutta RK3 was used in all the calculations. For each reconstruction algorithm, the first row displays the L1-norm errors for the $v_z$ component of the velocity after one period of time. The second row compiles the corresponding order of accuracy from contiguous spatial resolutions.}

\begin{tabular}[t]{lcccccccc}
\hline
\hline
\textbf{HLL}&N=8&N=16&N=32&N=64&N=128&N=256&N=512\\
\hline
Godunov&$0.242793$&$0.209234$&$0.157907$&$0.105586$&$0.063334$&$0.035223$&$0.018672$\\
&-&$0.214612$&$0.406045$&$0.580648$&$0.737373$&$0.846470$&$0.915669$\\
\hline
MinMod&$0.200698$&$0.100186$&$0.029124$&$0.006919$&$0.001937$&$0.000533$&$0.000139$\\
    &-&$1.002341$&$1.782403$&$2.073510$&$1.836458$&$1.861338$&$1.937243$\\
\hline    
MC&$0.148572$&$0.033643$&$0.006466$&$0.001386$&$0.000326$&$7.927\times10^{-5}$&$1.948\times10^{-5}$\\
    &-&$2.142766$&$2.379298$&$2.221720$&$2.086197$&$2.042077$&$2.024194$\\
\hline
VanLeer&$0.169355$&$0.050479$&$0.010169$&$0.001901$&$0.000398$&$9.017\times10^{-5}$&$2.112\times10^{-5}$\\
    &-&$1.746282$&$2.311483$&$2.418651$&$2.254662$&$2.143970$&$2.093994$\\
\hline
\end{tabular}

\label{table1}
\end{table*}



\begin{table*}
\centering
\caption{Same as Table \ref{table1} but employing the HLLC Riemann solver.}

\begin{tabular}[t]{lcccccccc}
\hline
\hline
\textbf{HLLC}&N=8&N=16&N=32&N=64&N=128&N=256&N=512\\
\hline
Godunov&$0.237358$&$0.192402$&$0.137008$&$0.087538$&$0.050748$&$0.027606$&$0.014446$\\
    &-&$0.302943$&$0.489863$&$0.646273$&$0.786574$&$0.878347$&$0.934353$\\
\hline
MinMod&$0.182472$&$0.081609$&$0.021935$&$0.005471$&$0.001513$&$0.000404$&$0.000104$\\
    &-&$1.160869$&$1.895440$&$2.003176$&$1.854302$&$1.902286$&$1.948340$\\
\hline    
MC&$0.126914$&$0.027667$&$0.005486$&$0.001279$&$0.000313$&$7.749\times10^{-5}$&$1.926\times10^{-5}$\\
    &-&$2.197577$&$2.334295$&$2.100195$&$2.030285$&$2.015140$&$2.007810$\\
\hline    
VanLeer&$0.148915$&$0.039794$&$0.007867$&$0.001579$&$0.000355$&$8.357\times10^{-5}$&$2.013\times10^{-5}$\\
    &-&$1.903846$&$2.338629$&$2.316817$&$2.151387$&$2.088442$&$2.053370$\\
\hline
\end{tabular}
\label{table2}
\end{table*}



\begin{table*}
\centering
\caption{Same as Table \ref{table1} but employing the HLLD Riemann solver and high-order spatial and temporal methods.}

\begin{tabular}[t]{lcccccccc}
\hline
\hline
\textbf{HLLD}&N=8&N=16&N=32&N=64&N=128&N=256&N=512\\
\hline
Godunov&$0.231373$&$0.182843$&$0.126489$&$0.078187$&$0.044176$&$0.023622$&$0.012235$\\
    &-&$0.339613$&$0.531601$&$0.694016$&$0.823657$&$0.903157$&$0.949035$\\
\hline
MinMod&$0.171364$&$0.073171$&$0.019213$&$0.004800$&$0.001308$&$0.000348$&$8.994\times10^{-5}$\\
    &-&$1.227719$&$1.929134$&$2.000842$&$1.875652$&$1.909334$&$1.953022$\\
\hline    
MC&$0.113706$&$0.025652$&$0.005504$&$0.001301$&$0.000316$&$7.790\times10^{-5}$&$1.932\times10^{-5}$\\
    &-&$2.148129$&$2.220420$&$2.080788$&$2.040914$&$2.020979$&$2.011657$\\
\hline    
VanLeer&$0.136352$&$0.036726$&$0.007871$&$0.001612$&$0.000360$&$8.445\times10^{-5}$&$2.026\times10^{-5}$\\
    &-&$1.892460$&$2.222038$&$2.287145$&$2.161129$&$2.094082$&$2.059075$\\
\hline    
MP3&$0.061922$&$0.009593$&$0.001285$&$0.000163$&$2.055\times10^{-5}$&$2.576\times10^{-6}$&$3.224\times10^{-7}$\\
    &-&$2.690331$&$2.899875$&$2.974505$&$2.991875$&$2.996154$&$2.998231$\\
\hline    
MP5&$0.009251$&$0.000425$&$2.651\times10^{-5}$&$2.456\times10^{-6}$&$2.813\times10^{-7}$&$3.440\times10^{-8}$&$4.284\times10^{-9}$\\
    &-&$4.443053$&$4.003761$&$3.432235$&$3.126066$&$3.031713$&$3.005290$\\
\hline    
MP5+RK4&$0.008389$&$0.000295$&$9.761\times10^{-6}$&$3.086\times10^{-7}$&$9.634\times10^{-9}$&$2.977\times10^{-10}$&$9.006\times10^{-12}$\\
    &-&$4.827224$&$4.920093$&$4.982834$&$5.001730$&$5.016265$&$5.046792$\\
    
\hline
\end{tabular}
\label{table3}
\end{table*}



\begin{table*}
\centering
\caption{Same as Table \ref{table1} but employing the PVM-int-8 Jacobian-free Riemann solver.}

\begin{tabular}[t]{lcccccccc}
\hline
\hline
\textbf{PVM}&N=8&N=16&N=32&N=64&N=128&N=256&N=512\\
\hline
Godunov&$0.236296$&$0.192590$&$0.136829$&$0.086507$&$0.049650$&$0.026810$&$0.013965$\\
    &-&$0.295066$&$0.493159$&$0.661477$&$0.801016$&$0.889058$&$0.940912$\\
\hline
MinMod&$0.181923$&$0.080879$&$0.021910$&$0.005388$&$0.001481$&$0.000401$&$0.000103$\\
    &-&$1.169493$&$1.884117$&$2.023693$&$1.863015$&$1.882146$&$1.950976$\\
\hline    
MC&$0.125110$&$0.028691$&$0.005910$&$0.001349$&$0.000322$&$7.873\times10^{-5}$&$1.942\times10^{-5}$\\
  &-&$2.124491$&$2.279383$&$2.130461$&$2.065503$&$2.034027$&$2.019233$\\
\hline    
VanLeer&$0.147824$&$0.040515$&$0.008698$&$0.001720$&$0.000375$&$8.675\times10^{-5}$&$2.061\times10^{-5}$\\
    &-&$1.867340$&$2.219613$&$2.337903$&$2.197052$&$2.112699$&$2.073304$\\
\hline 
MP3&$0.069036$&$0.011010$&$0.001475$&$0.000187$&$2.358\times10^{-5}$&$2.954\times10^{-6}$&$3.696\times10^{-7}$\\
    &-&$2.648510$&$2.899967$&$2.974409$&$2.992596$&$2.996515$&$2.998835$\\ 
\hline
\end{tabular}
\label{table4}
\end{table*}


\subsubsection{Riemann problem for the optically-thin radiation transport}
\label{rp-thin}

The second step in the validation of the new scheme consists in testing the explicit part of the radiation module, but neglecting any interaction between the fluid and the radiation field. For this purpose, we considered two Riemann problems proposed by \cite{pluto19} in the optically-thin regime.

\begin{figure*}
	\includegraphics[width=\textwidth]{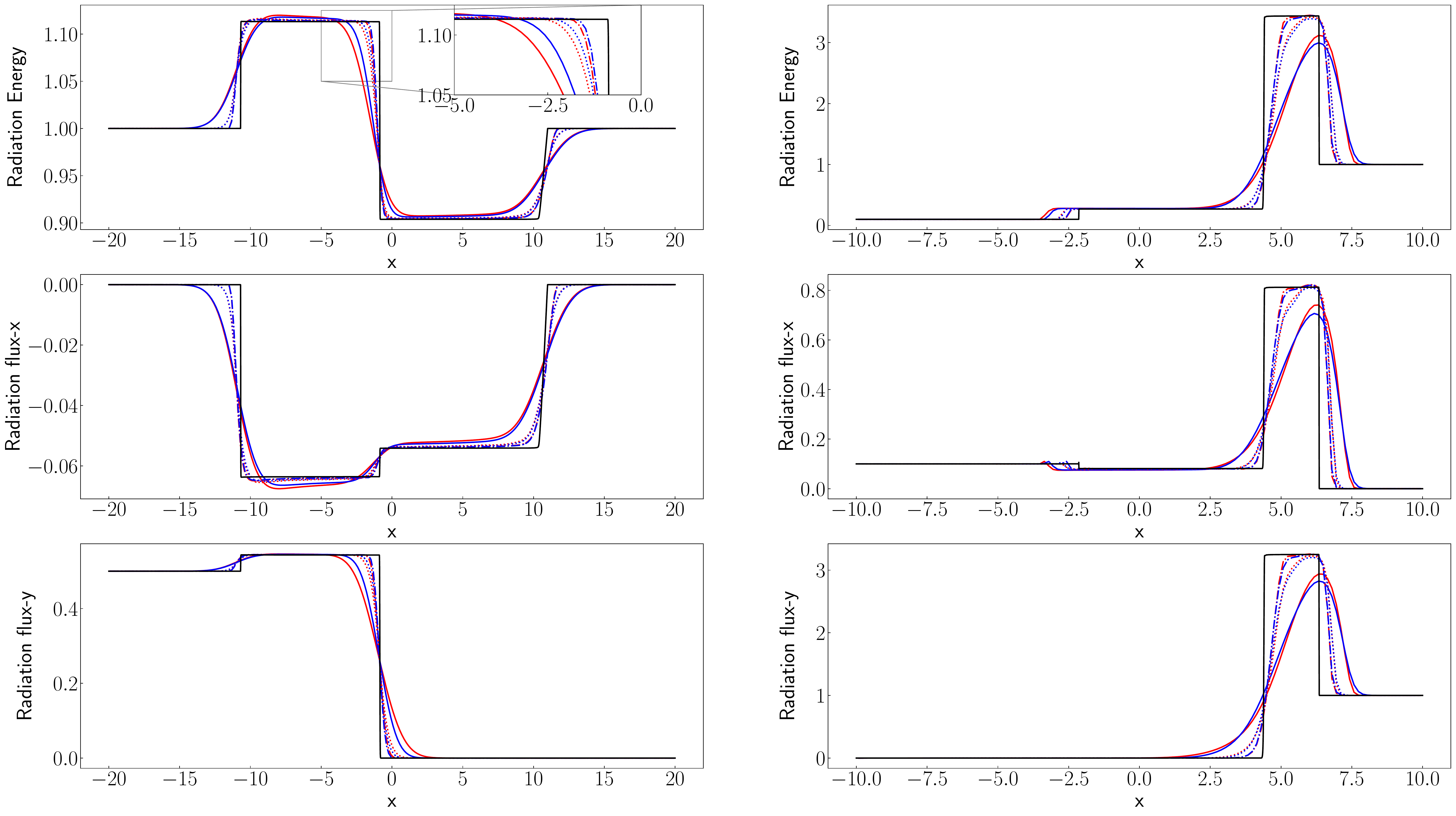}
    \caption{Radiation energy (top), radiation flux in the x-direction (middle) and radiation flux in the y-direction (bottom) for the optically-thin radiation transport problems Test 1 (left column) and Test 2 (right column). We show an approximation to the analytic solution at $t=20$ using $2^{14}$ zones (black solid line). The solution of the problem, computed with $2^8$ zones, is over-plotted for our two Riemann solvers: Rad-HLL (red) and PVM-int-8 (blue). We also compare the effect of different reconstructions: first-order (solid line), PLM-MinMod (dot line) and PLM-VanLeer (dash-dot line).}
    \label{thin}
\end{figure*}

The tests are initialized in a one-dimensional Cartesian grid that covered the space $[-20,20]$, where a discontinuity at $x=0$ separates two different radiation states. For Test 1, the radiation field is given by $(E_r, F_r^x, F_r^y)_L=(1,0,1/2)$ and $(E_r, F_r^x, F_r^y)_R=(1,0,0)$, while for Test 2, $(E_r, F_r^x, F_r^y)_L=(1/10,1/10,0)$ and $(E_r, F_r^x, F_r^y)_R=(1,0,1)$. In both cases, we run the test two times: first, we calculate an approximation to the analytic solution using high resolution ($2^{14}$ cells) and second-order reconstruction techniques (PLM-VanLeer). Then, we employed $2^8$ zones and first-order spatial reconstruction to solve the test with our two Riemann solvers: Rad-HLL and PVM-int-8. We considered outflow boundary conditions. Fig.~\ref{thin} shows the solution of these two problems at $t=20$. The pseudo-analytic solution is represented with a black solid line, while the results with $2^8$ zones and Rad-HLL and PVM-int-8 are shown in red and blue, respectively. The spatial reconstruction techniques are distinguished with a solid line (Godunov), dotted line (PLM-MinMod) and dash-dot line (PLM-VanLeer). 

As it is expected for the Jacobian decomposition of the Rad-RMHD equations (see Eqs.~\ref{jacobian}-\ref{lam3}), we obtained in both cases a three-wave pattern. In Test 1, we can distinguish a left-going shock, a right-going expansion wave and the analogous of a contact wave near $x=-1$ (Fig.~\ref{thin}, left column). In Test 2, we found a left-going shock, a right-going shock and a contact wave (Fig.~\ref{thin}, right column). Specially for Test 1, our implementation of PVM-int-8 gives a sharper solution around the contact wave than Rad-HLL. In Test 2, this discrepancy is mitigated. The effect of the Riemann solver on the solution accuracy is also less significant with the second order reconstruction, as it was previously mentioned in Sec.~\ref{dzaw}. 

\subsubsection{Shock tubes}

We consider a collection of four shock tube problems initially proposed in \cite{farris08}. These tests were solved by several authours using the \JLM{Eddington} approximation \citep{zanotti11,fragile12,taka13,sadowski13}, and later solved by \cite{mckinney14,rivera19,pluto19} for the M1 closure. These tests, which are similar to the Balsara shock tube problems of RMHD \citep{balsara01},  consist of two different radiation and fluid states separated by a discontinuity at $x=0$. The initial conditions of the problems are shown in the aforementioned references, so we do not reproduce them here. In all of these papers, the radiation field is given in the comoving frame, so we start by applying a Lorentz boost to convert all radiation quantities to the laboratory frame. This transformation is given by the following relations \citep[see e.g.,][]{park06,rivera19}:
\begin{equation}
    \tilde{E}_r=W^2\left(E_r-2v_i F_r^i + v_i v_j P^{i j}\right),
\end{equation}
\begin{equation}
\begin{aligned}
    \tilde{F}^{i}_r=-W^2v^{i}E+W\left[\delta_j^{i}+\left(\frac{W-1}{v^2}+W\right)v^{i}v_j\right]F^j_r-\\
    Wv_j\left(\delta_k^{i}+\frac{W-1}{v^2}v^iv_k\right)P^{jk},
\end{aligned}
\end{equation}
where replacing $\boldsymbol{v}$ for $-\boldsymbol{v}$ yields the inverse transformation.

The radiation constant for each problem, in the units of the code, is given in \cite{mckinney14}. We evolve the system until we achieve a final stationary state. For the first problem, the domain is the one-dimensional Cartesian grid $[-20,20]$ with a resolution of 800 zones. In all other cases, we consider the domain $[-40,40]$ with a resolution of 1600 zones to avoid that boundaries interfere in the stationarity of the final solution \citep{rivera19}. Outflow conditions are considered at both boundaries of the grid. We choose $\gamma=5/3$ for each test except for the highly-relativistic shock, where we choose $\gamma=2$. As in \cite{pluto19}, \JLM{initial} fluxes follows $\tilde{F}_r^x=0.01 \tilde{E}_r$.

\begin{figure}%
	\includegraphics[width=\linewidth]{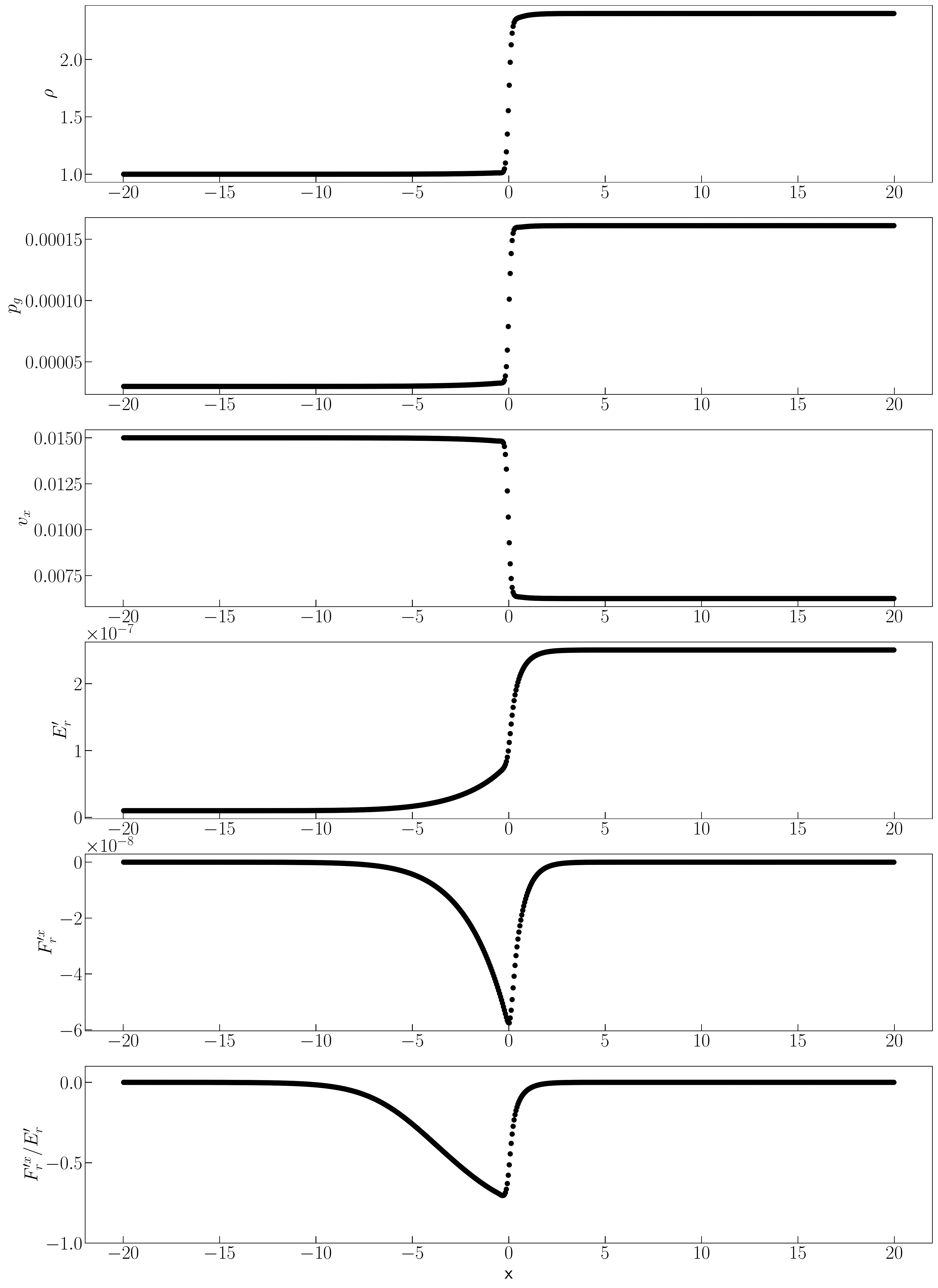}
   \caption{(a) Density, (b) gas pressure, (c) x-velocity, (d) radiation energy in the comoving frame, (e) radiation energy flux in the x direction and in the comoving frame and (f) reduced flux in the comoving frame for the non relativistic strong shock problem at $t=5000$, using 800 zones.}
    \label{sh1}
\end{figure}

\paragraph{Non-relativistic strong shock.} The initial conditions for this test are set such that the gas energy density dominates over the radiation energy density and the initial velocities are non-relativistic. The solution of the problem at $t=5000$ is shown in Fig~\ref{sh1}. Due to the initial conditions, radiation is transported from right to left, in front of the shock ($F_x<0$). Although fluid quantities show a strong discontinuity at $x = 0$, the radiation field is continuous in the whole domain. With the M1 closure, the radiation field decays with $e^{-\rho\kappa|x|}$ from the position of the shock \citep[see e.g.,][]{taka132}. Our solution is consistent with the results presented in the literature by other authors (both with the Eddington approximation and with M1, since for this test the results are almost similar) and with the semi-analytic solution \citep{farris08,fragile12,mckinney14}.

\begin{figure}%
	\includegraphics[width=\linewidth]{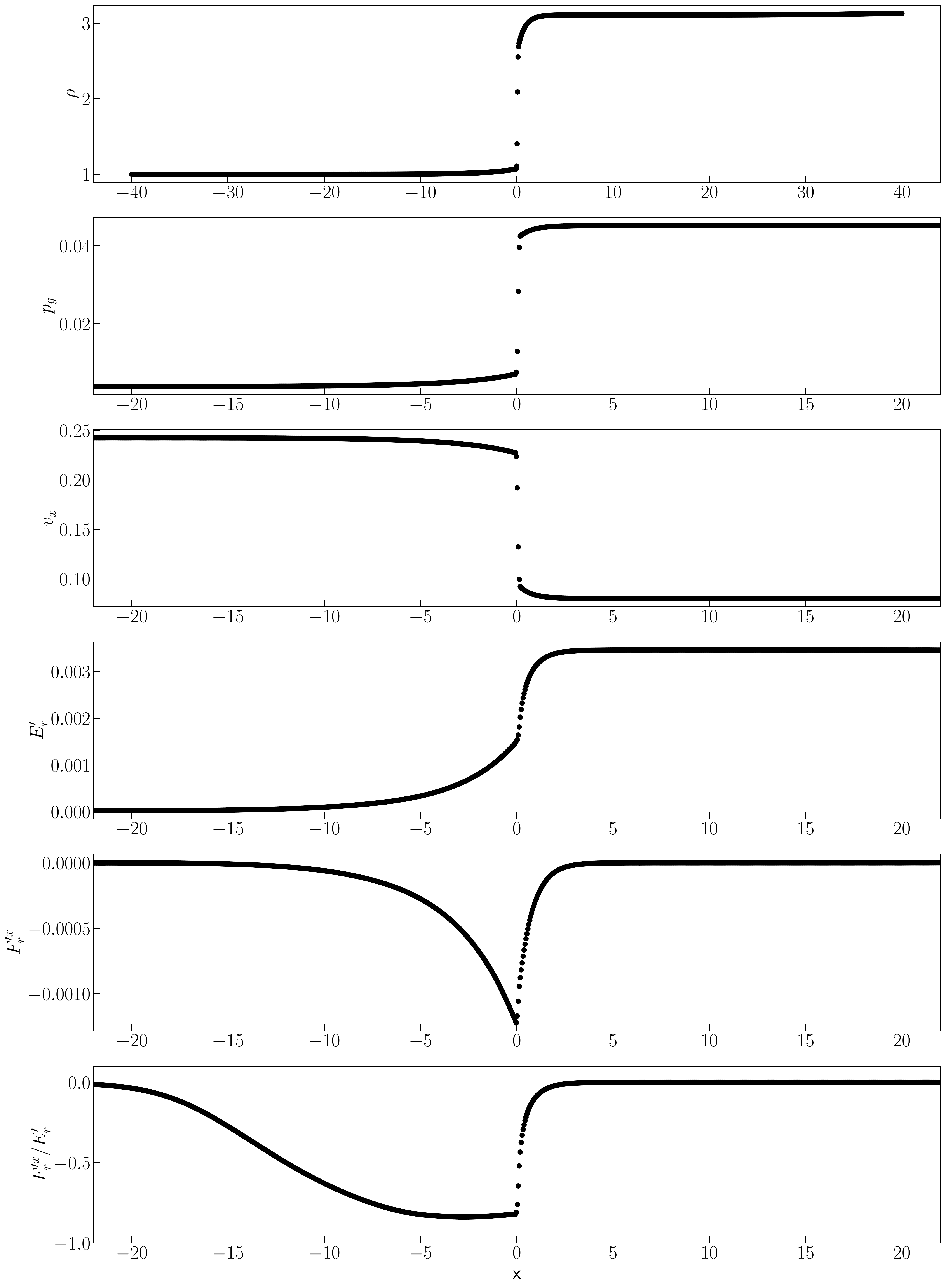}
   \caption{(a) Density, (b) gas pressure, (c) x-velocity, (d) radiation energy in the comoving frame, (e) radiation energy flux in the x direction and in the comoving frame and (f) reduced flux in the comoving frame for the mildly relativistic strong shock problem at $t=500$, using 1600 zones. Only a zoom of the [-20,20] central region is shown in the figure.}
    \label{sh2}
\end{figure}

\paragraph{Mildly-relativistic strong shock.} As in the non-relativistic strong shock, the gas energy density dominates over the radiation energy density, although in this case the initial velocities are higher ($u_L=0.25$). The solution of the problem at $t=500$ is shown in  Fig.~\ref{sh2}. Apart from the fact that the radiation energy profile is sharper near the shock, the solution is similar to the non-relativistic test. The M1 closure produces a smoothing of the numerical profiles at the shock position compared to the discontinuity found with the Eddington approximation \citep{farris08,taka13,tolstov15}.

\begin{figure}
	\includegraphics[width=\linewidth]{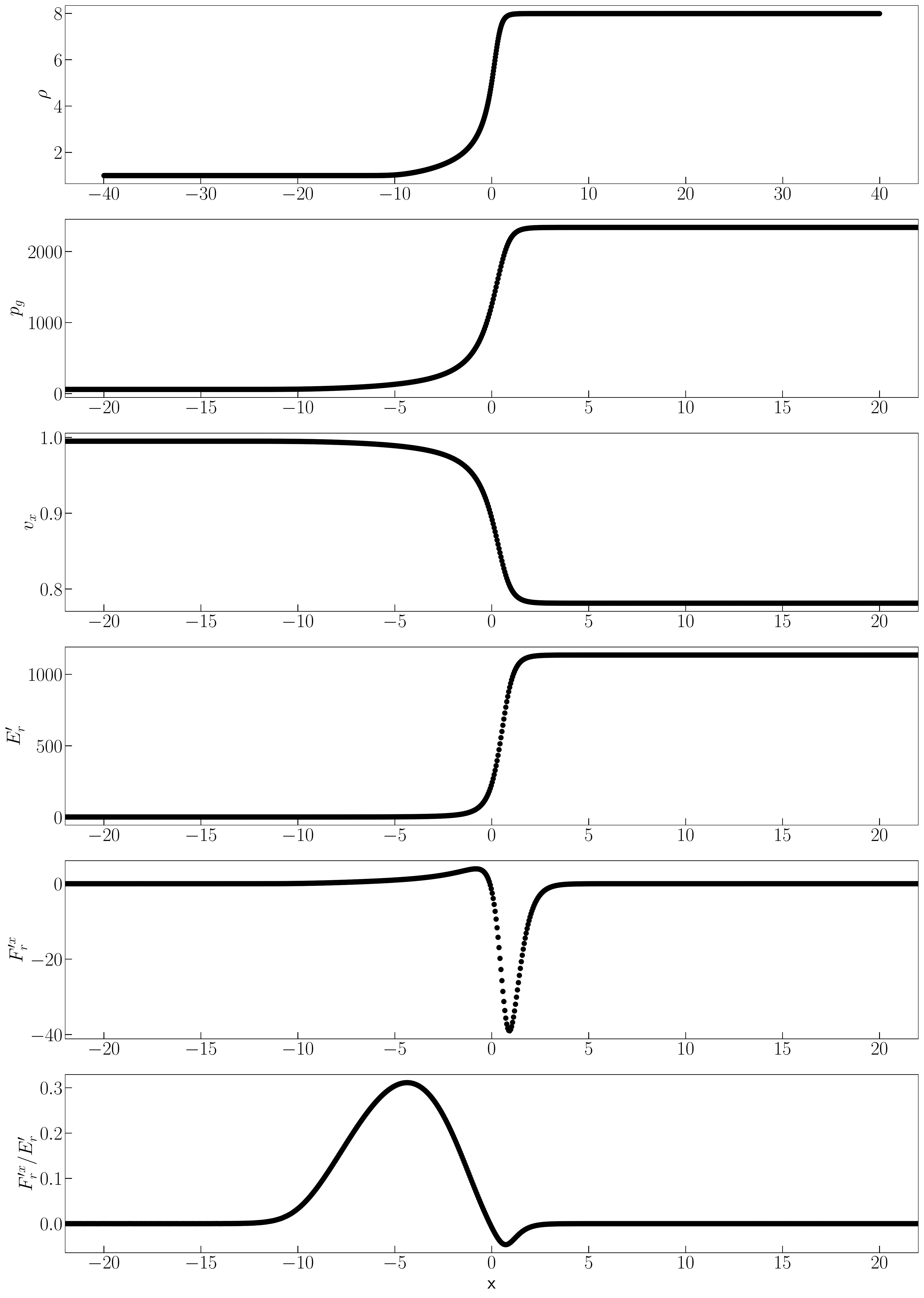}
   \caption{(a) Density, (b) gas pressure, (c) x-velocity, (d) radiation energy in the comoving frame, (e) radiation energy flux in the x direction and in the comoving frame and (f) reduced flux in the comoving frame for the highly relativistic strong shock problem at $t=500$, using 1600 zones. Only a zoom of the [-20,20] central region is shown in the figure.}
    \label{sh3}
\end{figure}

\paragraph{Highly-relativistic shock.} The initial conditions for this test are set such that the gas energy density dominates over the radiation energy density and the upstream Lorentz factor is highly relativistic, $W\approx 10$ \citep{farris08}. The solution of the problem at $t=500$ is shown in Fig.~\ref{sh3}. In this case, both the fluid quantities and the radiation field are smooth and continuous.
Although the shock front is stationary for the Eddington approximation, it drifts with a small velocity of $\sim 1.6\times 10^{-4}$ with the M1 closure \citep{taka132,rivera19}, which is not relevant for the time scales of the test. The maximum of $\tilde{F}_r^x/\tilde{E}_r$ is $\sim 0.31$, consistently with the literature \citep[see e.g.,][]{taka132}.

\begin{figure}
	\includegraphics[width=\linewidth]{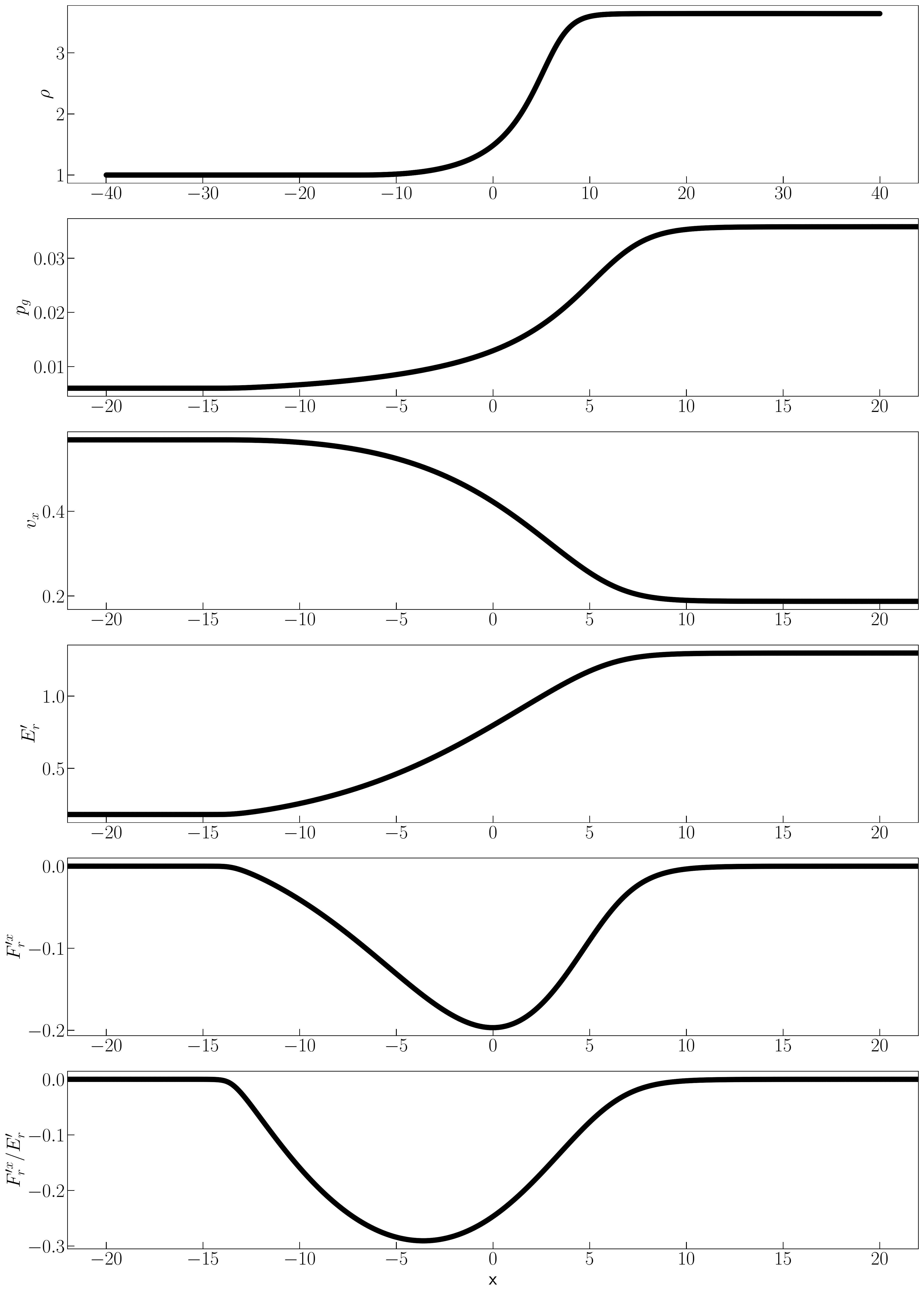}
   \caption{(a) Density, (b) gas pressure, (c) x-velocity, (d) radiation energy in the comoving frame, (e) radiation energy flux in the x direction and in the comoving frame and (f) reduced flux in the comoving frame for the non relativistic strong shock problem at $t=500$, using 1600 zones. Only a zoom of the [-20,20] central region is shown in the figure.}
    \label{sh4}
\end{figure}

\paragraph{Radiation-pressure-dominated shock.} In this problem, the initial conditions are set such that the radiation pressure dominates over the gas pressure and the upstream velocity is mildly relativistic. The solution of the problem at $t=500$ is shown in Fig~\ref{sh4}. After the initial discontinuity at $x=0$ breaks up, fluid and radiation achieve a steady state solution with a small drift velocity \citep{taka132}, where the shock front remains approximately at the origin. As in the previous tests, radiation is transported from the left to the right ($\tilde{F}_r^x<0$), penetrating up to $x\sim -12$, while the radiation flux is reduced by absorption. Since matter and radiation are strongly coupled, radiation force produces a decrease in the fluid velocity, while density is enhanced. This solution is smooth and continuous for both fluid and radiation quantities, and it can be compared with the results shown in the literature elsewhere \citep{taka132,pluto19}.

\subsubsection{Optically-thick radiation pulse}
\label{sec:pulse}

\begin{figure*}
	\includegraphics[width=\textwidth]{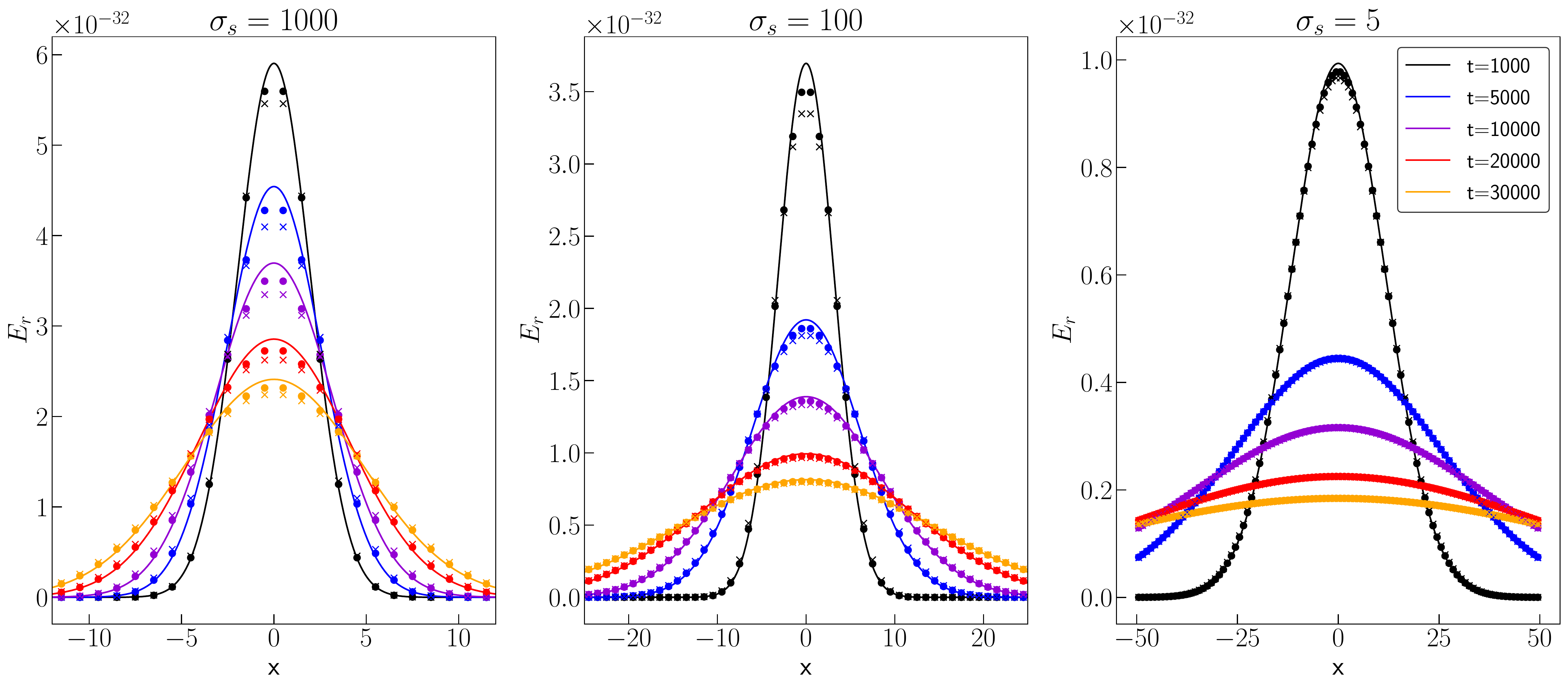}
    \caption{Radiation energy density for the optically-thick radiative pulse diffusion at $\sigma_s=1000$ (left), $\sigma_s=100$ (middle) and $\sigma_s=5$ (right). The analytic solution (solid lines) is over-plotted with the numerical results with the PVM-int-8 Riemann solver (dots) and the Rad-HLL Riemann solver (crosses) for different time frames.}
  \label{gaus}
\end{figure*}

The last one-dimensional problem of this section consists on a radiative pulse propagating along one of the spatial Cartesian coordinates in an optically-thick medium \citep{mckinney14,rivera19,pluto19}. Assuming initial thermal equilibrium, the radiation energy profile is given by:
\begin{equation}
    E_r=a_r\left[T_0\left(1+100~e^{-x^2/\omega^2}\right)\right]^{4},
    \label{gaus_init}
\end{equation}
where $T_0=10^6$ and $\omega=5$. We choose the value of the radiation constant in the code units as $a_r=6.24\times 10^{-64}$ such that the ratio $k=T\rho/p=1$. The background density is $\rho=1$ and the adiabatic exponent, $\gamma=5/3$. With these initial conditions, the evolution of the system follows \citep{mckinney14}:
\begin{equation}
    E_r(t)=A~\mathrm{exp}\left(\frac{-x^2}{4D~(t+t_0)}\right)\left(\frac{t+t_0}{t_0}\right)^{-1/4},
    \label{gaus_analytic}
\end{equation}
where $A\approx 6.49\times 10^{-32}$, $t_0\approx 4800$, and $D=1/(3(\kappa+\sigma_s))$. For this test, the absorption opacity is set to zero while we run the pulse for three different values of the scattering coefficient, {namely $\sigma_s = 5, 100, 1000$}. The grid covers the domain $[-50,50]$ and we use 100 computation zones. With the chosen values of the grid size, the corresponding cell opacities for the runs are $\tau = 5, 100, 1000$. To avoid the effects of the grid boundaries, the analytic solution is imposed there. Fig.~\ref{gaus} shows the solution of the problem at different times, compared to the analytic solution given by Eq.~\ref{gaus_analytic}. Dots represent the solution with the PVM-int-8 Riemann solver and crosses show the results with Rad-HLL, where the maximum and minimum velocities were limited according to Eq.~\ref{limits}. As suggested by \cite{mckinney14}, the solution of the central points exhibits more diffusion at earlier times, while the effect is mitigated later in the evolution. The similarity of our results with respect to the analytic solution for different values of the scattering opacity demonstrates a good performance of our scheme in the optically thick regime, where PVM-int-8 improves slightly the accuracy of Rad-HLL. On the other hand, the solution of the problem with zeroth-order reconstruction, PVM-int-8 with $\beta=1$ or Rad-HLL with no velocity limiting, results in catastrophic diffusive solutions (specially for $\tau>100$).

\subsection{Two and three dimensions}
\label{2-3d}

We propose a collection of multidimensional problems to test the performance of the new implemented methods in more than one spatial dimension. In order to avoid spurious oscillations, we introduced a shock flattening around strong shocks and sharp radiation gradients \citep{mignone05}. In particular, the VanLeer reconstruction algorithm is degraded to the more diffusive MinMod algorithm around strong shocks.

\subsubsection{Cylindrical/spherical magnetized blast wave}

\begin{figure}%
	\includegraphics[width=\linewidth]{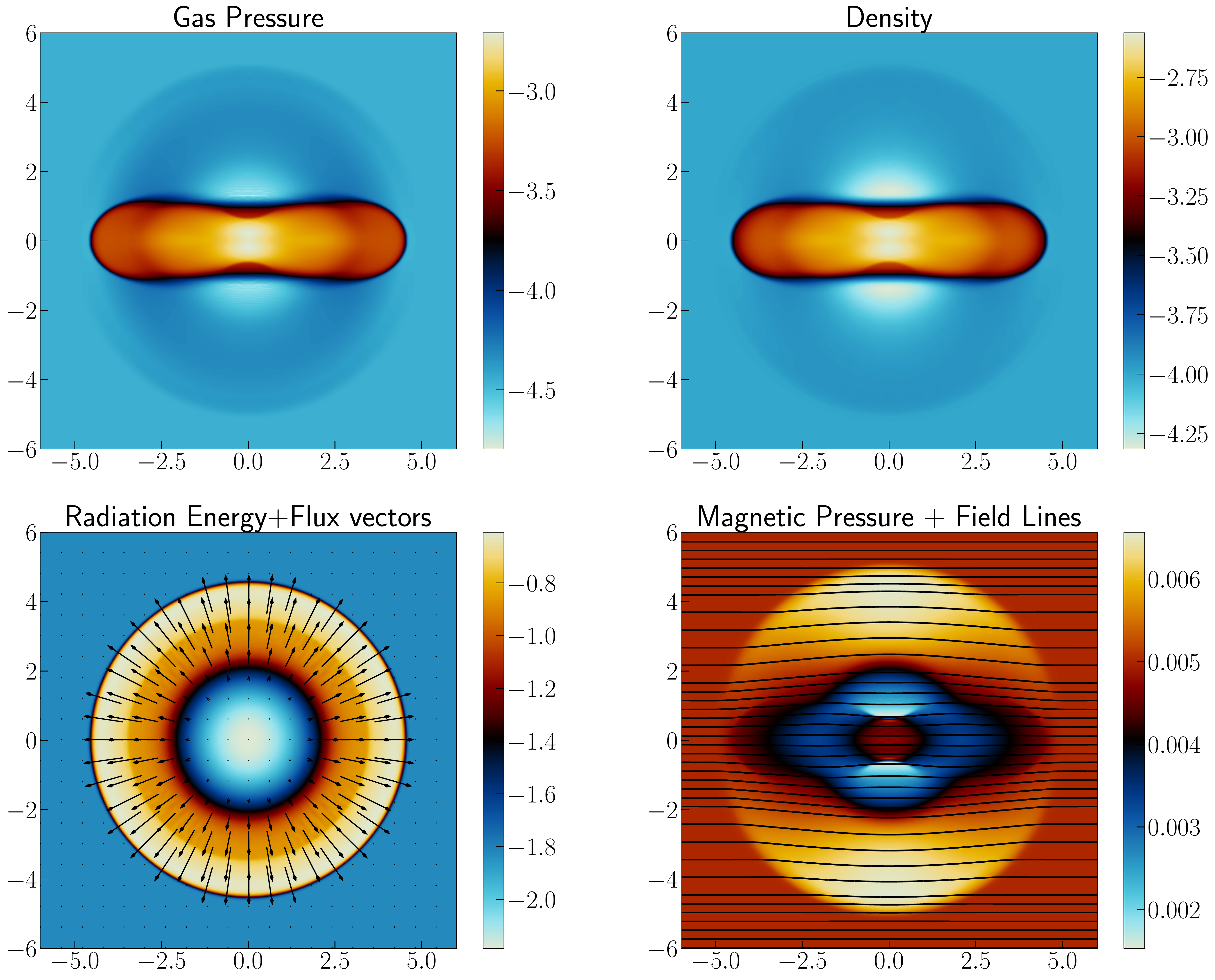}
    \caption{Logarithmic gas pressure (top left), logarithmic density (top right), logarithmic radiation energy density (bottom left) and logarithmic magnetic pressure (bottom right) at $t=4$ for the two-dimensional cylindrical magnetized blast wave. In the bottom left panel, the vector field represents the radiation flux, which propagates radially from the center. Magnetic field lines are superposed to the magnetic pressure. We consider the Cartesian grid [-6,6]$\times$[-6,6] with $360$ cells per spatial dimension. The absorption coefficient is $\kappa=1$.}
  \label{cmbw}
\end{figure}

\begin{figure}%
	\includegraphics[width=\linewidth]{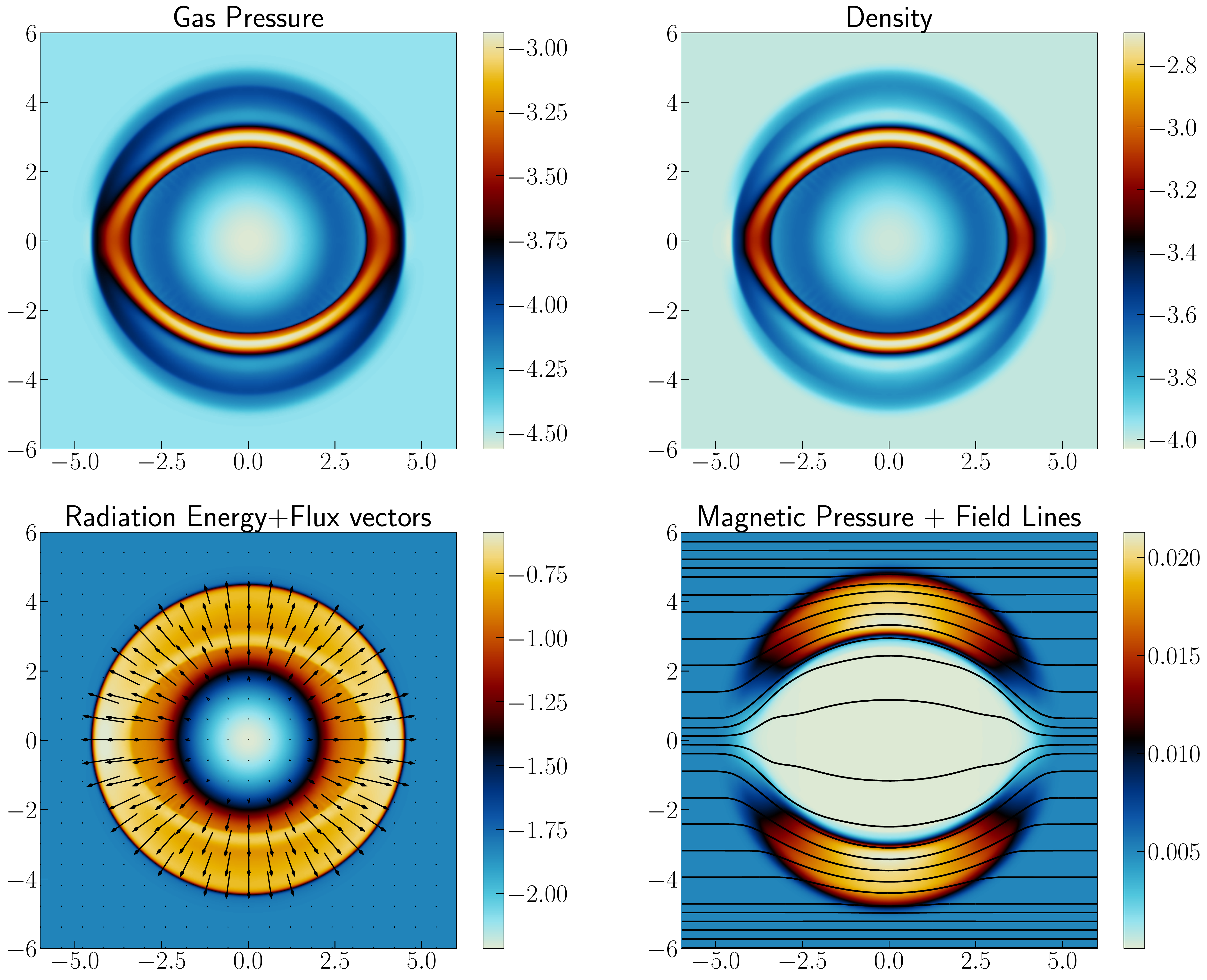}
    \caption{Same as Fig.~\ref{cmbw} for an absorption coefficient $\kappa=1000$.}
  \label{cmbw2}
\end{figure}

\begin{figure}%
\begin{center}
	\includegraphics[width=\linewidth]{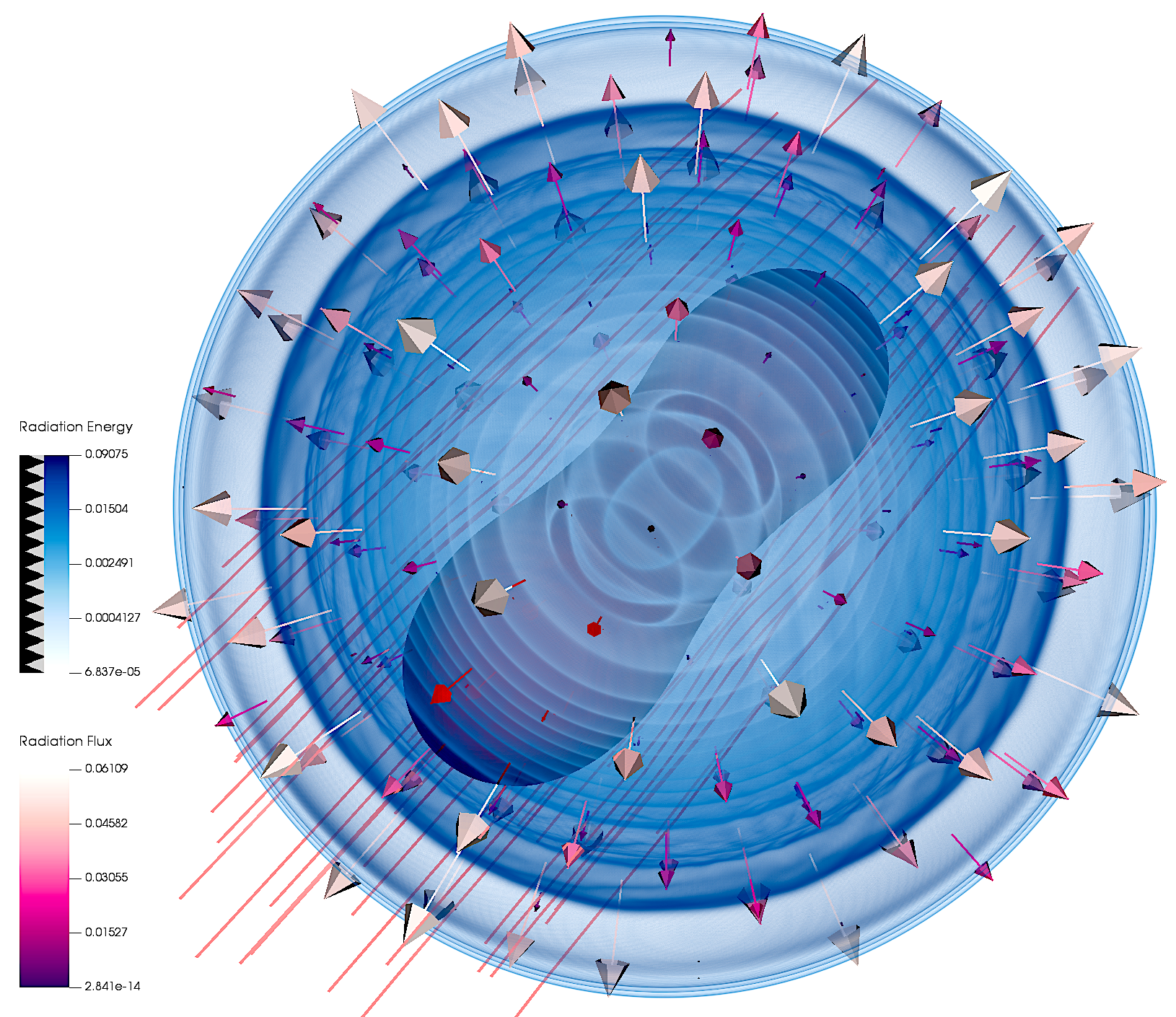}
    \caption{Radiation energy density for the optically-thin, spherical magnetized blast wave at $t=4$. The vector field represents the radiation flux. Magnetic field lines (pink solid lines) are slightly deformed by the flow.}
  \label{smbw}
\end{center}
\end{figure}

The cylindrical/spherical magnetized blast wave is a classical problem in RMHD and has been extensively used to test the performance of the HRSC numerical methods handling MHD wave degeneracies parallel and perpendicular to the field orientation \citep[see e.g.,][and references therein]{martiRev}. In LM22, we solved the test in RMHD for both weak and strong magnetization regimes. The version that we consider in this paper was first proposed in \cite{pluto19} and it is the first and only test that we solve considering both radiation and magnetic fields. However, unlike \cite{pluto19}, we also consider the spherical version of the magnetized blast wave in three dimensions.
The main particularity of this test in radiation magnetohydrodynamics is that it can switch from the radiation-dominated to the magnetically dominated regime varying the medium's opacity; when the opacity is small, the flow is magnetically dominated, but radiation dominates the blast wave dynamics in the optically-thick regime. We consider the two-dimensional computational grid $[-6,6]\times[-6,6]$ with $360^2$ cells, such that the initial geometry is identical to LM22. The ambient density and pressure are $\rho_0=10^{-4}$, $p_0=3.49\times 10^{-5}$, while inside the cylinder/sphere of unit radius they become $\rho_1=10^{-2}$, $p_1=1.31\times 10^{-2}$. The adiabatic exponent is $\gamma=4/3$. The magnetic field is aligned with the x-direction, $\boldsymbol{B}=(0.1,0,0)$, and the fluid is initially at rest (i.e., $\boldsymbol{v}=0$). For the radiation field, we assume thermal equilibrium with the plasma and $\boldsymbol{F}_r=0$. The opacity of the medium is controlled with the absorption coefficient, which is $\kappa=1$ for the magnetically dominated problem and $\kappa=1000$ for the radiation dominated explosion, while the scattering opacity is $\sigma_s=0$ in the two cases. 
For $\kappa = 1$, the initial optical depth along the central sphere is $\tau_1 = 2 \rho_1 \kappa = 0.02 \ll 1$ (optically thin case). For $\kappa = 1000$, $\tau_1 = 20$. Fig.~\ref{cmbw} and Fig.~\ref{cmbw2} show the solution of the problem at $t=4$ for $\kappa=1$ and $\kappa=1000$, respectively. In the optically-thin scenario, flow dynamics is barely affected by the presence of radiation and the evolution is dominated by the magnetic field. Thus, flow quantities like pressure or density develop an elongated horizontal structure, following the magnetic field lines, which are slightly deformed by the wave, as in the RMHD case (see e.g., LM22). The maximum Lorentz factor that we get is $W_{\rm max}\approx 1.7$, which is similar to \cite{pluto19}. Since photons can diffuse freely, radiation energy density preserves the cylindrical symmetry. However, in the optically-thick problem ($\kappa=1000$), radiation pressure is dominant and matter is strongly coupled to the photon field. In this case, the flow recover an oblated ring shape, with $W_{\rm max}\approx 2.72$ along the x-axis. Magnetic field lines are highly deformed and the radiation energy density is no longer symmetric because of the interaction with the gas. Due to the exigent test conditions (source terms are very \textit{stiff}), in this latter problem we used the Rad-HLL Riemann solver, the MinMod slope limiter and a CFL=0.01 to avoid the appearance of non-physical solutions during the simulation. Fig.~\ref{smbw} shows the solution at $t=4$ of a three dimensional version of the optically-thin explosion (numerical resolution: 256 computational zones per spatial dimension). Since the symmetry of the radiation field is well preserved, we demonstrated the capability of our scheme to handle both radiation and magnetic fields in full three dimensional applications.

\subsubsection{The shadow problem}
\label{sec:shadow}

\begin{figure}%
	\includegraphics[width=\linewidth]{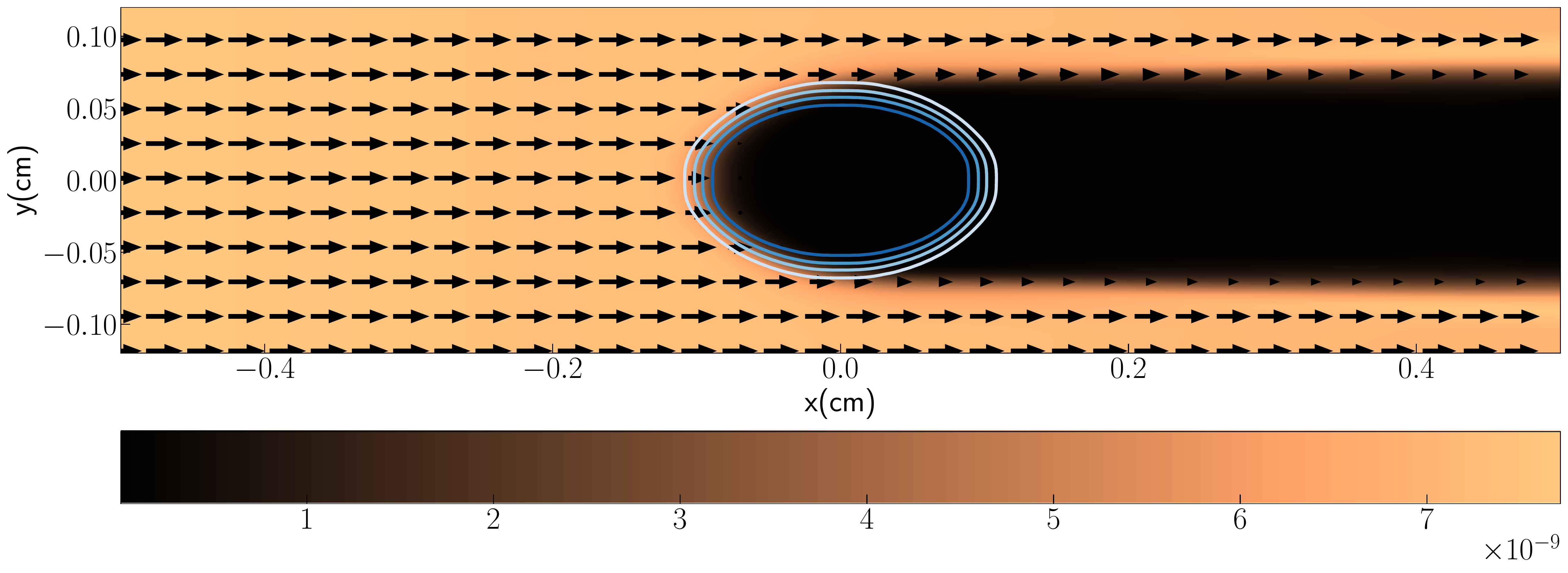}
    \caption{Radiation energy density for the single shadow problem at $t=1.50\,t_c$. The vector field represents the radiation flux. Contours show the position of the optically-thick blob.}
 \label{shadow}
\end{figure}

\begin{figure*}%
	\includegraphics[width=\linewidth]{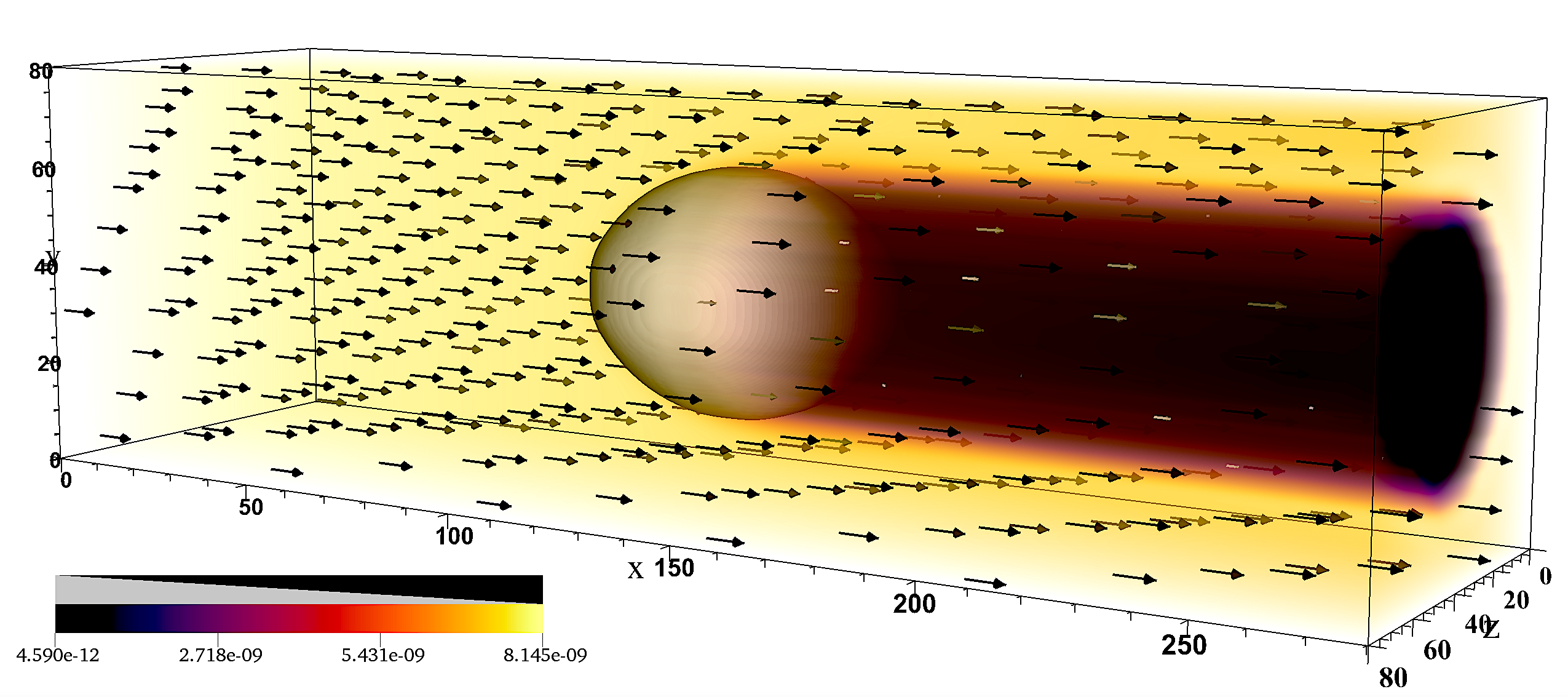}
    \caption{Radiation energy density for the single shadow problem in three dimensions at $t=1.50\,t_c$. The vector field represents the radiation flux. The white contour shows the position of the optically-thick blob.}
 \label{shadow3D}
\end{figure*}

\paragraph{Single shadow.} The shadow test is one of the most common problems in radiation hydrodynamics to test the performance of the M1 closure \citep{hayes03,gonzalez07,sadowski13,mckinney14,fragile14,rivera19,pluto19,weih20}. This problem consists on a free-streaming radiation beam that impacts into a highly opaque region, producing a shadow behind it. Although there are multiple versions of the shadow problem, in this paper we follow the initial conditions given by \cite{gonzalez07,fragile14} or \cite{pluto19}. We consider the two dimensional Cartesian grid $[-0.5,0.5]\times[-0.12,0.12]$~cm, with a resolution of $280\times 80$ computational cells. The initial density distribution is given by:
\begin{equation}
    \rho(x,y)=\rho_0+\frac{\rho_1-\rho_0}{1+e^{\Delta}},
\end{equation}
\begin{equation}
    \Delta=10\left[\left(\frac{x}{x_0}\right)^2+\left(\frac{y}{y_0}\right)^2-1\right],
\end{equation}
where $\rho_0=1$~g/cm$^3$, $\rho_1=10^3$~g/cm$^3$ and $(x_0,y_0)=(0.10,0.06)$~cm. Matter and radiation are initially in thermal equilibrium with a temperature $T=290$~K and we consider constant absorption opacity $\kappa=0.1$~cm$^2$/g, while we neglect photon scattering. 
Fluxes are set to zero and matter is also initially at rest. From the left boundary (at $x=-0.5$~cm), we inject a radiation beam at temperature $T=1740$~K and flux $\boldsymbol{F}_r=(cE_r,0,0)$ (i.e., radiation is in the free streaming limit). In the rest of the boundaries, we impose outflow conditions. The adiabatic exponent is $\gamma=5/3$. Fig.~\ref{shadow} shows the solution of the problem at $t=1.5\,t_c$, where $t_c = 1$ cm/$c$. The highly opaque ellipsoid (represented with contours) produces a shadow behind it that can be compared with the aforementioned references. A three-dimensional version of this problem, simulated with the same initial conditions, is shown in Fig.~\ref{shadow3D} at $t=1.5\,t_c$. For this favourable setup, the M1 closure favours that the flux remains parallel to the injection direction casting a shadow behind the 3D ellipsoid.

\begin{figure}%
\includegraphics[width=\linewidth]{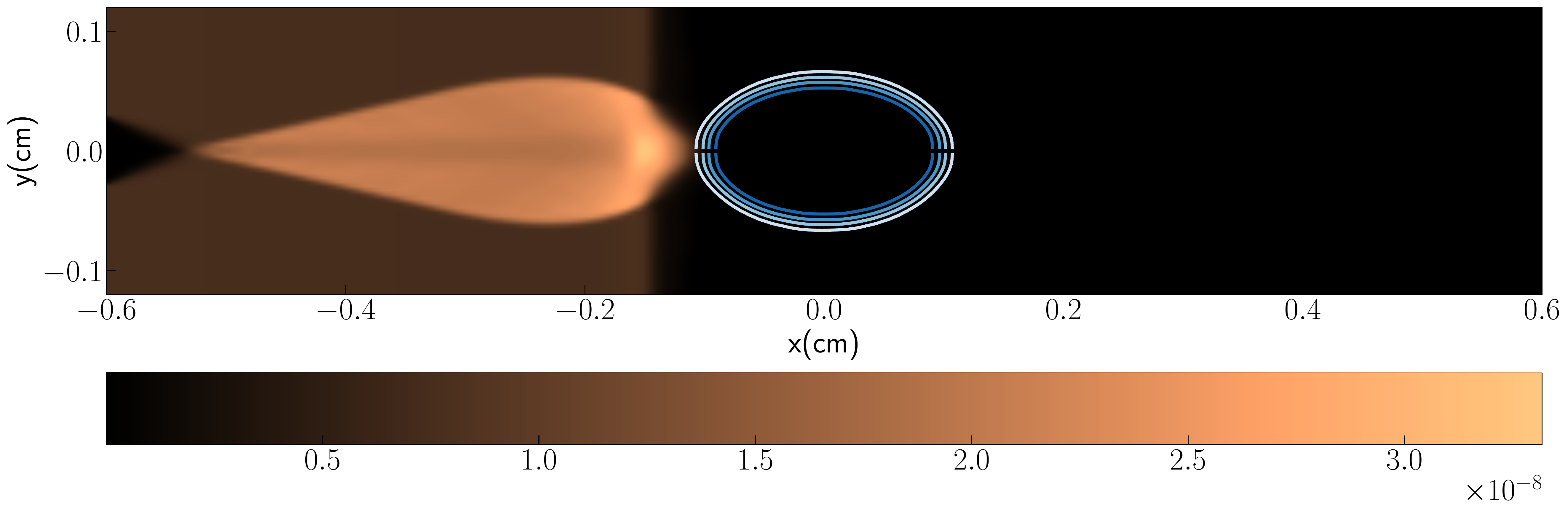}
\includegraphics[width=\linewidth]{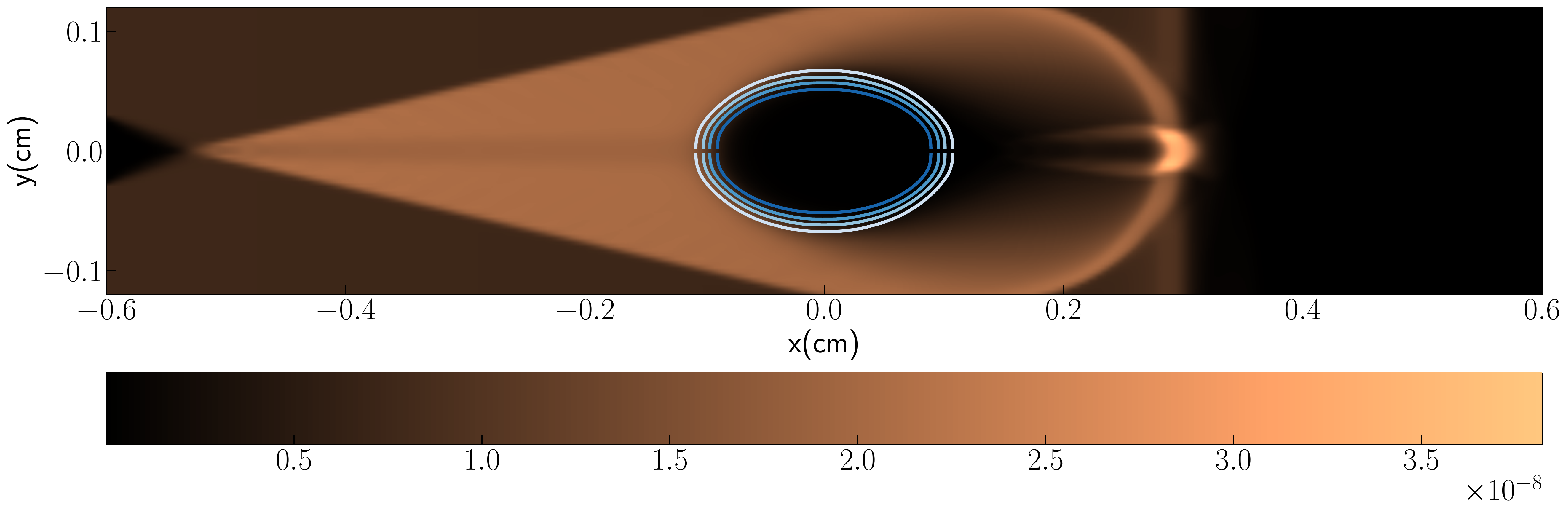}
\includegraphics[width=\linewidth]{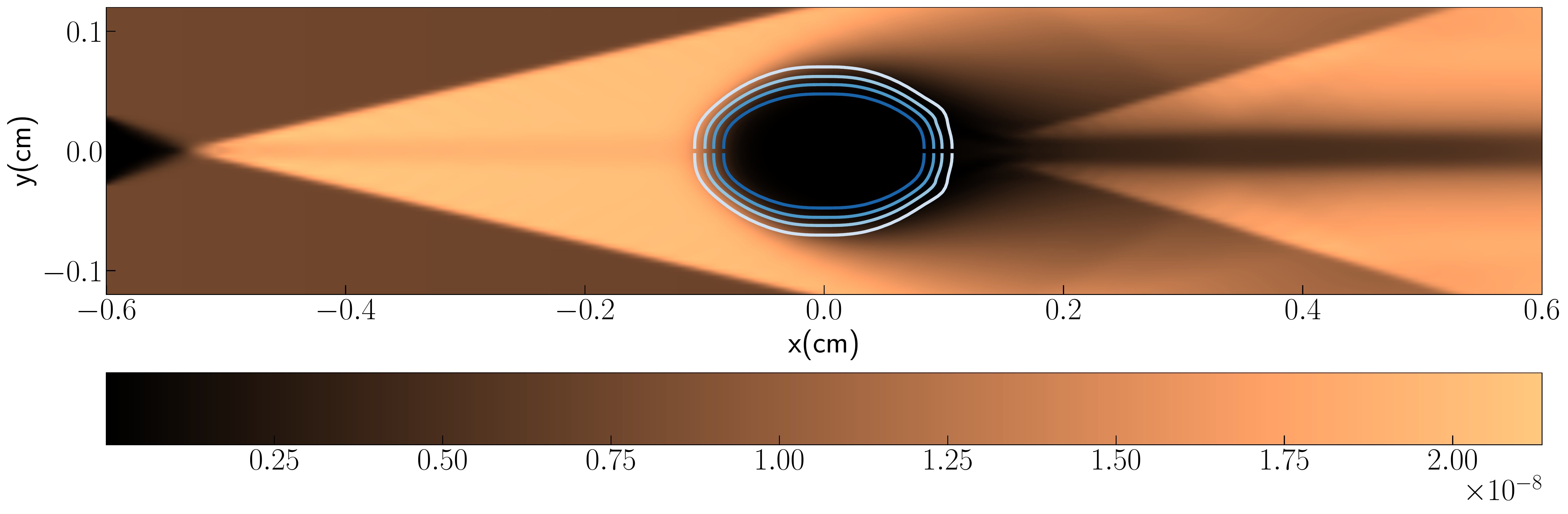}
    \caption{Radiation energy density for the single shadow problem at $t=0.50\,t_c$ (top panel), $t=1.00\,t_c$ (middle panel) and $t=4.50\,t_c$ (bottom panel). Contours show the position of the optically-thick blob.}
 \label{adow2}
\end{figure}

\paragraph{Double shadow.} We consider an extension of the shadow problem that we described in the previous paragraph to the case of multiple sources of light, where the M1 closure is expected to be less accurate \citep{sadowski13,mckinney14,rivera16,rivera19}. The initial setup is the same as in the single shadow test, although in this case we consider a slightly bigger domain: $[-0.6,0.6]\times[0,0.15]$~cm, with the same cell size than before. For this problem, we only simulate the upper half of the grid since we establish reflecting symmetry boundary conditions at $y=0$. In the left x-boundary (only where $y>0.03$~cm), we inject an inclined radiation beam with the same radiation energy than in the single shadow problem, but in this case the radiation flux is given by $F_r^x=0.93 c E_r$ and $F_r^y=-0.37 c E_r$. Due to the reflecting boundary at $y=0$, the optically-thick ellipsoid is effectively illuminated by two self-crossing beams of light. In the remaining boundaries, we assume outflow conditions. In the bottom panel of Fig.~\ref{adow2}, we show the solution of the problem at $t=4.5\,t_c$. In the left x-boundary, we can distinguish a small triangular region which is not illuminated. After this region, approximately at $x\approx-0.45$~cm, radiation overlaps and the radiative energy density increases as $E_r=2E_r^0$, while the radiation flux aligns with the x-direction with $F_r^x=0.93 c E_r$. Therefore, since the radiation flux is no longer $F_r^x\approx c E_r$ as in the single shadow problem, the implied distribution of specific intensity is not pointing along the x-axis but spreads in other directions about this axis \citep[see the discussion in ][]{sadowski13}. As a result, the radiation front is indeed an elongated ellipsoid pointing along the x-direction (Fig.~\ref{adow2}, top panel).

Eventually, this has an effect on the shadow characteristics behind the clump, with a region of total shadow (umbra) limited by the edges of a wider region of partial shadow (penumbra). As it is evident from the bottom panel of Fig.~\ref{adow2}, there is also a narrow horizontal shadow extended along the x-axis. This non-physical feature, which has been shown in the literature \citep[see e.g.,][and references therein]{sadowski13,mckinney14,rivera19}, is an artefact related with the M1 closure.

\section{Discussion and conclusions}
\label{sec5}

In this paper, we have presented a novel scheme for treating radiation transport within \textsc{Lóstrego}, a fully parallel 3D RMHD code presented for the first time in LM22.

In our approach, the equations of relativistic radiation magnetohydrodynamics, which are formulated taking successive moments of the Boltzmann radiative transfer equation, are solved under the gray-body approximation and the M1 closure using an IMEX time integration scheme. This closure scheme allows to handle accurately both the free-streaming and diffusion radiation transport limits. We have introduced, for the first time in the context of Rad-RMHD, a new family of approximate Riemann solvers based on internal polynomial approximations to the viscosity matrix (PVM methods; see Sect.~\ref{PVM}). The underlying polynomia of this group of solvers admit a Jacobian-free formulation of the scheme, where only evaluations of the conserved variables and flux vectors are required. In the context of Rad-RMHD, this is particularly useful because the characteristic speeds of the magnetofluid and the radiation subsystem of equations can be -in general- very different, and therefore these two blocks must be independently handled when using Riemann solvers based on characteristic wave decomposition to avoid excessive diffusion. Moreover, the maximum and minimum characteristic speeds of the radiation block must be limited in the optically-thick regime for the same reason. As we described in the paper, the PVM-int solver overcomes these issues in a very natural and elegant way. On the one hand, the same algorithm is used to compute the numerical fluxes of the full Rad-RMHD system of equations. On the other hand, the numerical viscosity of the original solver is reduced gradually as the opacity of the numerical cells around an interface increases.

We tested the new PVM-int Riemann solver in the context of RMHD, comparing its performance with the HLL family. For that purpose, we also introduced in \textsc{Lóstrego} high-order methods both in space and time, although in this paper we kept second-order nominal accuracy (PLM+RK3) to test the new radiation module. For the cell reconstruction step, we have introduced the MP scheme of \cite{suresh97}, which yields from third to ninth spatial order of accuracy. The algorithm was tested with a smooth RMHD test problem that consists on the propagation of a large-amplitude, circularly polarized Alfvén wave along a uniform background magnetic field (see Sect.~\ref{dzaw}). We demonstrated that the third order MP reconstruction (i.e., MP3) together with RK3 (MP3+RK3), yields third order of accuracy for all of our Riemann solvers, including the new PVM-int-8. This extends the results of \cite{marti152}, where the code was tested up to second order. \cite{castro17} also showed third order of accuracy with the PVM family of Riemann solvers, but using the Piecewise Hyperbolic Method \citep[PHM;][]{marquina94}. Nevertheless, when we increased the order of spatial reconstruction (using, e.g., MP5), the order of accuracy of the code was limited to the order of the Runge-Kutta (in this case, RK3), although the L1 errors were almost two orders of magnitude lower than with MP3+RK3 (see Table~\ref{table3}). Thus, MP5+RK3 is a good configuration for applications involving smooth solutions that required high degree of accuracy (for example, development of instabilities or perturbation analysis in RMHD). To achieve an order of accuracy higher than third order, we have implemented a fourth-order (RK4), strong-stability-preserving RK scheme, based on five integration steps \citep{balsara17}. Using this algorithm together with MP5 (MP5+RK4) we were able to achieve the nominal fifth order of accuracy, for which we used the HLLD Riemann solver because it was the least diffusive scheme in the benchmark.

Let us note that \cite{delzanna07} probed fifth order of accuracy by limiting \textit{adhoc} the integration time step, but this approach is unpractical for large-scale simulations. The same applies to the RK4 algorithm of \cite{balsara01}, which is computationally expensive due to the large number of RK substeps. A specific benchmark of these high-order methods based on real multi-dimensional (Rad)-RMHD applications will be addressed in future work and presented elsewhere.

Focusing on Rad-RMHD, we have used a collection of one-dimensional and multi-dimensional test problems (see Sects.~\ref{rp-thin}~-~\ref{sec:pulse} and \ref{2-3d}) to demonstrate that our scheme is robust and accurate for systems with different initial conditions, including both scattering and absorption opacity in the free-streaming and diffusion radiation transport limits. The convergence analysis presented on Tables~\ref{table1}-\ref{table4} and displayed in Fig.~\ref{L1} shows that our PVM-int-8 solver performs similar to HLLC, although all solvers behaved similar with first order linear reconstruction, as expected. In radiation magnetohydrodynamics, PVM-int-8 improves the accuracy of Rad-HLL, specially around strong discontinuties of the radiation field (see Fig.~\ref{thin}). Comparing with \cite{pluto19}, the difference between these two solvers seems to be less prominent than the one reported for Rad-HLL and the new radiation version of HLLC. However, in the optically-thick regime PVM-int-8 improves slightly the accuracy of Rad-HLL, specially in those regions that show more numerical diffusion (see central region of the gaussian pulse in Fig.~\ref{gaus}), where Rad-HLLC is not valid. This is the first time to our knowledge that a Riemann solver improves the accuracy of Rad-HLL in the highly optically thick regime ($\tau \gg 1$). Moreover, we want to stress out that our choice of the parameter $\beta$ (Eq.~\ref{beta}), devised to reduce the numerical diffusion in the optically thick regime, is somehow arbitrary. 

As previously mentioned, our definition of $\beta$ relies on its simplicity and the results in one-dimensional tests, but other functions should be explored in the future.
Similarly, other schemes of the Jacobian-free family should be also adapted and tested for Rad-RMHD. For example, \cite{castro17} found that the PVM-Cheb-12 \cite[based on the Chebyshev approximation of degree 12;][]{castro16}, or the DOT-Cheb-12 and DOT-int-8 approximate DOT solvers using the same polynomials but a Gauss-Legendre quadrature \citep{castro16}, gave promising results in one-dimensional RMHD applications. Indeed, PVM-Cheb-12 and DOT-Cheb-12 are slightly less difussive than PVM-int-8, although the latter automatically satisfies the stability condition needed to ensure the robustness and convergence of the schemes. \JLM{Moreover, \cite{castro17} also showed that, at least for the one-dimensional Brio-Wu test problem, PVM-based methods were a better option than DOT-based ones from the point of view of computational efficiency, and that using PVM-int-8 increases the total CPU time with respect to HLL by a factor of $\sim 2$. This is mainly because PVM requires multiple calls to the recovery of primitive variables, whose algorithm is computationally expensive. Since the variables of the radiation field are at the same time primitive and conserved, the additional cost of our version of PVM with respect to Rad-HLL will not be affected by this, but only by the number of additional operations and flux evaluations, which will also increase for Rad-HLL.}
In two-dimensions, all schemes were similar, but DOT solvers are \JLM{also} computationally more expensive. We also tested in \textsc{Lóstrego} the PVM-int-16 solver, but the results were identical to PVM-int-8 at higher computational cost, so we did not reproduce the results in this paper. In conclusion, the Jacobian-free character of this type of solvers (1) allows to solve the equations of Rad-RMHD with the same algorithm without decomposing the system in two independent blocks, (2) improves the accuracy of Rad-HLL both in the optically thin and optically-thick radiation transport limits and (3) it is robust and accurate for several test problems in one and more than one spatial dimensions. 

We have also shown that the M1 closure is also capable to reproduce shadows in two and three dimensions, although it is less accurate when multiple sources of light are involved in optically-thin regions, leading to numerical artefacts and non-physical solutions (as shown in the double-shadow test problem of Sec.~\ref{sec:shadow}). In order to treat accurately this scenario, other authors have proposed -in the context of radiation hydrodynamics and magnetohydrodynamics- to compute a closure relation directly from the Eddington tensor \citep[i.e., the variable Eddington tensor formalism,][]{stone92,gonzalez07,jiang12}. A deep revision of the closure scheme will be, however, addressed in future work.

\section*{Ackowledgements}
The project that gave rise to these results received the support of a fellowship from ”La Caixa” Foundation (ID 100010434). The fellowship code is LCF/BQ/DR19/11740030. J.L.M acknowledges additional support from the Spanish Ministerio de Ciencia through grant PID2019-105510GB-C31/AEI/10.13039/501100011033.
M.P. and J.M.-M acknowledge support by the Spanish Ministerio de Ciencia through grants PID2019-107427GB-C33 and PGC2018-095984-B-I00, and from the Generalitat Valenciana through grant PROMETEU/2019/071. M.P acknowledges additional support from the Spanish Ministerio de Ciencia through grant $\mathrm{PID}2019-105510\mathrm{GB}-\mathrm{C}31/\mathrm{AEI}/10.13039/501100011033.$
Computer simulations have been carried out  in the Servei d'Inform\`atica de la Universitat de Val\`encia (Lluis Vives cluster). We thank the referees for all the constructive comments and suggestions that helped to improve the quality of the manuscript. We also thank David Melon Fuksman, Manuel J. Castro, José M. Gallardo and Martin Obergaulinger for useful comments and discussion during the implementation of the algorithm.

\section*{Data Availability}
The algorithms that support the results of this paper and the current version of our code \textsc{Lóstrego} are available from the corresponding author upon reasonable request.





\bibliographystyle{elsarticle-harv}
\bibliography{bib}







\end{document}